\input harvmac
%\draftmode

\input amssym

\def\gt{\tilde{g}}

\baselineskip 13pt

%\CveticUJ
\lref\CveticUJ{
  M.~Cveti\v c and D.~Youm,
``Dyonic BPS saturated black holes of heterotic string on a six torus,''
Phys.\ Rev.\ D {\bf 53}, 584 (1996).
[hep-th/9507090].
%%CITATION = hep-th/9507090%%
}

%\CveticBJ
\lref\CveticBJ{
  M.~Cveti\v c and A.~A.~Tseytlin,
  ``Solitonic strings and BPS saturated dyonic black holes,''
Phys.\ Rev.\ D {\bf 53}, 5619 (1996), [Erratum-ibid.\ D {\bf 55}, 3907 (1997)].
[hep-th/9512031].
%%CITATION = hep-th/9512031%%
}
%\ChongNA
\lref\ChongNA{
  Z.~-W.~Chong, M.~Cveti\v c, H.~L\"u and C.~N.~Pope,
  %``Charged rotating black holes in four-dimensional gauged and ungauged supergravities,''
Nucl.\ Phys.\ B {\bf 717}, 246 (2005).
[hep-th/0411045].
%%CITATION = hep-th/0411045%%
}

%\ChakrabortyNU
\lref\ChakrabortyNU{
  A. Chakraborty and C. Krishnan,
``Subttractors,''
JHEP {\bf 1308}, 057 (2013).  [arXiv: 1212.1875 [hep-th]].
%%CITATION = arXiv:1212.1875%%
}

%\ChakrabortyFX
\lref\ChakrabortyFX{
  A.~Chakraborty and C.~Krishnan,
``Attraction, with Boundaries,''
Class.\ Quant.\ Grav.\  {\bf 31}, 045009 (2014).
[arXiv:1212.6919 [hep-th]].
%%CITATION = arXiv:1212.6919%%
}

%\HorowitzAC
\lref\HorowitzAC{
  G.~T.~Horowitz, D.~A.~Lowe and J.~M.~Maldacena,
  ``Statistical entropy of nonextremal four-dimensional black holes and U duality,''
Phys.\ Rev.\ Lett.\  {\bf 77}, 430 (1996).
[hep-th/9603195].
%%CITATION = hep-th/9603195%%
}

%\CveticDN
\lref\CveticDN{
  M.~Cveti\v c and  F.~Larsen,
  ``Conformal Symmetry for Black Holes in Four Dimensions,''
JHEP {\bf 1209}, 076 (2012).
[arXiv:1112.4846 [hep-th]].
%%CITATION = arXiv:1112.4846%%
}

%\ChowTIA
\lref\ChowTIA{
  D.~D.~K.~Chow and G.~Comp{\`e}re,
  ``Seed for general rotating non-extremal black holes of N=8 supergravity,''
Class.\ Quant.\ Grav.\  {\bf 31}, 022001 (2014).
[arXiv:1310.1925 [hep-th]].
%%CITATION = arXiv:1310.1925%%
}

%\LarsenDH
\lref\LarsenDH{
  F.~Larsen and E.~J.~Martinec,
  ``Currents and moduli in the (4,0) theory,''
JHEP {\bf 9911}, 002 (1999).
[hep-th/9909088].
%%CITATION = hep-th/9909088%%
}

%\LarsenQR
\lref\LarsenQR{
  F.~Larsen,
  ``Entropy of thermally excited black rings,''
JHEP {\bf 0510}, 100 (2005).
[hep-th/0505152].
%%CITATION = hep-th/0505152%%
}

%\ChowCCA
\lref\ChowCCA{
  D.~D.~K.~Chow and G.~Comp{\`e}re,
  ``Black holes in N=8 supergravity from SO(4,4) hidden symmetries,''
[arXiv:1404.2602 [hep-th]].
%%CITATION = arXiv:1404.2602%%
}

%\CveticZQ
\lref\CveticZQ{
  M.~Cveti\v c and C.~M.~Hull,
  ``Black holes and U duality,''
Nucl.\ Phys.\ B {\bf 480}, 296 (1996).
[hep-th/9606193].
%%CITATION = hep-th/9606193%%
}

%\SenEB
\lref\SenEB{
  A.~Sen,
  ``Black hole solutions in heterotic string theory on a torus,''
Nucl.\ Phys.\ B {\bf 440}, 421 (1995).
[hep-th/9411187].
%%CITATION = hep-th/9411187%%
}

%\LarsenPU
\lref\LarsenPU{
  F.~Larsen,
  ``Kaluza-Klein black holes in string theory,''
[hep-th/0002166].
%%CITATION = hep-th/0002166%%
}

%\LarsenPP
\lref\LarsenPP{
  F.~Larsen,
  ``Rotating Kaluza-Klein black holes,''
Nucl.\ Phys.\ B {\bf 575}, 211 (2000).
[hep-th/9909102].
%%CITATION = hep-th/9909102%%
}

%\DijkgraafIT
\lref\DijkgraafIT{
  R.~Dijkgraaf, E.~P.~Verlinde and H.~L.~Verlinde,
  ``Counting dyons in N=4 string theory,''
Nucl.\ Phys.\ B {\bf 484}, 543 (1997).
[hep-th/9607026].
%%CITATION = hep-th/9607026%%
}

%\RasheedZV
\lref\RasheedZV{
  D.~Rasheed,
  ``The Rotating dyonic black holes of Kaluza-Klein theory,''
Nucl.\ Phys.\ B {\bf 454}, 379 (1995).
[hep-th/9505038].
%%CITATION = hep-th/9505038%%
}

%\CveticHP
\lref\CveticHP{
  M.~Cveti\v c and F.~Larsen,
  ``Conformal Symmetry for General Black Holes,''
JHEP {\bf 1202}, 122 (2012).
[arXiv:1106.3341 [hep-th]].
%%CITATION = arXiv:1106.3341%%
}

%\BaggioDB
\lref\BaggioDB{
  M.~Baggio, J.~de Boer, J.~I.~Jottar and D.~R.~Mayerson,
  ``Conformal Symmetry for Black Holes in Four Dimensions and Irrelevant Deformations,''
JHEP {\bf 1304}, 084 (2013).
[arXiv:1210.7695 [hep-th]].
%%CITATION = arXiv:1210.7695%%
}

%\CastroFD
\lref\CastroFD{
  A.~Castro, A.~Maloney and A.~Strominger,
  ``Hidden Conformal Symmetry of the Kerr Black Hole,''
Phys.\ Rev.\ D {\bf 82}, 024008 (2010).
[arXiv:1004.0996 [hep-th]].
%%CITATION = arXiv:1004.0996%%
}

%\CveticXV
\lref\CveticXV{
  M.~Cveti\v c and F.~Larsen,
  ``Grey body factors for rotating black holes in four-dimensions,''
Nucl.\ Phys.\ B {\bf 506}, 107 (1997).
[hep-th/9706071].
%%CITATION = hep-th/9706071%%
}

%\DabholkarZZ
\lref\DabholkarZZ{
  A.~Dabholkar and S.~Nampuri,
  %``Quantum black holes,''
Lect.\ Notes Phys.\  {\bf 851}, 165 (2012).
[arXiv:1208.4814 [hep-th]].
%%CITATION = arXiv:1208.4814%%
}

\lref\KrausWN{
  P.~Kraus,
  ``Lectures on black holes and the AdS(3) / CFT(2) correspondence,''
Lect.\ Notes Phys.\  {\bf 755}, 193 (2008).
[hep-th/0609074].
%%CITATION = hep-th/0609074%%
}

%\CveticKV
\lref\CveticKV{
  M.~Cveti\v c and D.~Youm,
  ``All the static spherically symmetric black holes of heterotic string on a six torus,''
Nucl.\ Phys.\ B {\bf 472}, 249 (1996).
[hep-th/9512127].
%%CITATION = hep-th/9512127%%
}

%\CveticKVa
\lref\CveticKVa{
  M.~Cveti\v c and D.~Youm,
  ``Entropy of nonextreme charged rotating black holes in string theory,''
Phys.\ Rev.\ D {\bf 54}, 2612 (1996).
[hep-th/9603147].
%%CITATION = hep-th/9603147%%
}

%\CveticVP
\lref\CveticVP{
  M.~Cveti\v c and F.~Larsen,
  ``Black hole horizons and the thermodynamics of strings,''
Nucl.\ Phys.\ Proc.\ Suppl.\  {\bf 62}, 443 (1998), [Nucl.\ Phys.\ Proc.\ Suppl.\  {\bf 68}, 55 (1998)].
[hep-th/9708090].
%%CITATION = hep-th/9708090%%
}

%\CveticCJA
\lref\CveticCJA{
  M.~Cveti\v c, M.~Guica and Z.~H.~Saleem,
``General black holes, untwisted,''
JHEP {\bf 1309}, 017 (2013).
[arXiv:1302.7032 [hep-th]].
%%CITATION = arXiv:1302.7032%%
}

%\CveticINA
\lref\CveticINA{
  M.~Cveti\v c, G.~W.~Gibbons and Z.~H.~Saleem,
``Quasi-Normal Modes for Subtracted Rotating and Magnetised Geometries,''
[arXiv:1401.0544 [hep-th]].
%%CITATION = UPR-1258-T%%
}

%\CveticLFA
\lref\CveticLFA{
  M.~Cveti\v c and G.~W.~Gibbons,
``Exact quasi-normal modes for the near horizon Kerr metric,''
Phys.\ Rev.\ D {\bf 89}, 064057 (2014).
[arXiv:1312.2250 [gr-qc]].
%%CITATION = UPR-1257-T%%
}

%\CveticTR
\lref\CveticTR{
  M.~Cveti\v c and G.~W.~Gibbons,
``Conformal Symmetry of a Black Hole as a Scaling Limit: A Black Hole in an Asymptotically Conical Box,''
JHEP {\bf 1207}, 014 (2012).
[arXiv:1201.0601 [hep-th]].
%%CITATION = arXiv:1201.0601%%
}

%\ShihUC
\lref\ShihUC{
  D.~Shih, A.~Strominger and X.~Yin,
  ``Recounting Dyons in N=4 string theory,''
JHEP {\bf 0610}, 087 (2006).
[hep-th/0505094].
%%CITATION = hep-th/0505094%%
}

%\VirmaniKW
\lref\VirmaniKW{
  A.~Virmani,
  ``Subtracted Geometry From Harrison Transformations,''
JHEP {\bf 1207}, 086 (2012).
[arXiv:1203.5088 [hep-th]].
%%CITATION = AEI-2012-031%%
}

%\CveticUW
\lref\CveticUW{
  M.~Cveti\v c and F.~Larsen,
  ``General rotating black holes in string theory: Grey body factors and event horizons,''
Phys.\ Rev.\ D {\bf 56}, 4994 (1997).
[hep-th/9705192].
%%CITATION = hep-th/9705192%%
}

%\MaldacenaDE
\lref\MaldacenaDE{
  J.~M.~Maldacena, A.~Strominger and E.~Witten,
  ``Black hole entropy in M theory,''
JHEP {\bf 9712}, 002 (1997).
[hep-th/9711053].
%%CITATION = hep-th/9711053%%
}

%\BehrndtJN
\lref\BehrndtJN{
  K.~Behrndt, G.~Lopes Cardoso, B.~de Wit, R.~Kallosh, D.~Lust and T.~Mohaupt,
  ``Classical and quantum N=2 supersymmetric black holes,''
Nucl.\ Phys.\ B {\bf 488}, 236 (1997).
[hep-th/9610105].
%%CITATION = hep-th/9610105%%
}

%\ShmakovaNZ
\lref\ShmakovaNZ{
  M.~Shmakova,
  ``Calabi-Yau black holes,''
Phys.\ Rev.\ D {\bf 56}, 540 (1997).
[hep-th/9612076].
%%CITATION = hep-th/9612076%%
}

%\BehrndtHU
\lref\BehrndtHU{
  K.~Behrndt, R.~Kallosh, J.~Rahmfeld, M.~Shmakova and W.~K.~Wong,
  ``STU black holes and string triality,''
Phys.\ Rev.\ D {\bf 54}, 6293 (1996).
[hep-th/9608059].
%%CITATION = SLAC-PUB-9856%%
}

%\SenEB
\lref\SenEB{
  A.~Sen,
  ``Black hole solutions in heterotic string theory on a torus,''
Nucl.\ Phys.\ B {\bf 440}, 421 (1995).
[hep-th/9411187].
%%CITATION = hep-th/9411187%%
}

%\BreckenridgeIS
\lref\BreckenridgeIS{
  J.~C.~Breckenridge, R.~C.~Myers, A.~W.~Peet and C.~Vafa,
  ``D-branes and spinning black holes,''
Phys.\ Lett.\ B {\bf 391}, 93 (1997).
[hep-th/9602065].
%%CITATION = hep-th/9602065%%
}

\Title{\vbox{\baselineskip12pt 
\hbox{UPR-1260-T} 
\vskip-.5in}
}
{\vbox{\centerline {Black Holes with Intrinsic Spin}} } 

\centerline{Mirjam Cveti\v{c}\foot{cvetic@hep.upenn.edu}$^{,3}$ and Finn
Larsen\foot{larsenf@umich.edu}}

\bigskip
\centerline{${}^1$\it{Department of Physics and Astronomy,
University of Pennsylvania,}}\centerline{\it{ Philadelphia, PA 19104,
USA.}}\vskip.2cm \centerline{${}^2$\it{Michigan Center for
Theoretical Physics,}} \centerline{\it{University of Michigan, Ann
Arbor, MI 48109-1120, USA.}}\vskip.2cm \centerline{${}^3$\it{Center for Applied Mathematics and Theoretical Physics,
}} \centerline{\it{University of Maribor, Maribor, Slovenia.}}

\baselineskip15pt

\vskip .3in

\centerline{\bf Abstract}
We analyze the general black hole solutions to the four dimensional STU model recently  constructed by Chow and Comp{\`e}re. We define a dilute gas 
limit where the black holes can be interpreted as excited states of an extremal ground state. In this limit we express the black hole entropy 
and the excitation energy in terms of physical quantities with no need for parametric charges. We discuss a dual microscopic CFT 
description that incorporates all electric and magnetic charges. This description is recovered geometrically by identification of a near horizon BTZ region. We construct the subtracted geometry with no restrictions on charges by analyzing the scalar wave equation in the full geometry. We determine the matter sources that support the subtracted geometry by studying a scaling limit and show that the general geometry permits a dilute gas description with parameters that we specify. 

%%%
\Date{June, 2014}

\newsec{Introduction}
In general relativity black holes are uniquely specified by their asymptotic quantum numbers: mass $M$, angular momentum $J$, and $U(1)$ charges $Q_i, P_i$. Theories arising from string theory at low energy typically feature many $U(1)$ charges so the most general solutions in these theories involve many parameters and the explicit solutions can be unwieldy. It is therefore preferable to consider only representative subsets of solutions. In the context of string theory black holes in four dimensions one of the most common settings for research in the area are the ``four charge" solutions to ${\cal N}=2$ supergravity with STU pre-potential \refs{\CveticUJ,\CveticKVa,\HorowitzAC,\BehrndtHU}. In the extremal limit these black holes are essentially interpreted as marginal bound states of four distinct types of branes. The ``four charge" family is readily embedded into important theories such as ${\cal N}=4$ and ${\cal N}=8$ supergravity and it is commonly considered representative of more general settings. 

The standard ``four charge'' solutions can be made more representative upon action by dualities. Solutions generated this way have essentially the same physical properties as their seeds, at least classically. However, the ``four charge'' family does {\it not} realize the most general charge vectors through this generating mechanism: there is precisely {\it one} parameter that is missing \refs{\SenEB,\CveticUJ}. This feature is special to four dimensions where it holds in the STU-model, in ${\cal N}=4,8$ SUGRA, and in other settings as well \CveticZQ. The special nature of the ``fifth charge" is due to the interactions of electric and magnetic fields with respect to the {\it same} $U(1)$ gauge field. In this case the crossing of fields famously gives rise to an angular momentum proportional to $\vec{Q}\cdot \vec{P}$. The spin parameter proportional to $\vec{Q}\cdot \vec{P}$ is composed of charges so it is {\it independent} of the angular momentum $J$ encoded in the asymptotic metric. We refer to this parameter as the {\it intrinsic spin} of the black hole. 

The central motivation for studying black holes in string theory is the prospect of a microscopic understanding of their internal structure and, in particular, the black hole entropy. The recent review \DabholkarZZ\ presents the state of the art. The intrinsic spin $\vec{Q}\cdot \vec{P}$ of the black hole is important to microscopic considerations already in the BPS setting that is so well studied (e.g., \refs{\DijkgraafIT,\ShihUC}).\foot{For an early study of the  microscopic origin of the intrinsic spin, in the BPS setting of the NS-NS sector, see \CveticBJ.} The interplay between intrinsic spin $\vec{Q}\cdot \vec{P}$ and external angular momentum $J$ is an essential aspect of efforts to understand black holes away from extremality. 

Unfortunately the physics of the two types of spin has turned out to be very complex already at the level of classical solutions. Early efforts constructed rotating black hole solutions with general charges except for the internal spin \refs{\CveticKVa,\ChongNA} and these efforts independently constructed solutions with the internal spin incorporated but no overall angular momentum \refs{\CveticBJ,\CveticKV}. These constructions employed the solution generating techniques, by acting  with a subset of O(4,4) transformations on a neutral  (Kerr or Schwarzschild)  black hole, reduced on time-like Killing vector to three dimensions.  Only recently did  a remarkable {\it tour de force} by  Chow and Comp{\`e}re  lead to a construction  of the most general black hole solutions in the STU model \refs{\ChowTIA,\ChowCCA}.  While employing the same solutions generating techniques, the missing parameter way obtained by acting with symmetry transformations  on a neutral Taub-NUT solution.  The physics of these new solutions is far from self-evident. The geometries are exceptionally involved and the physical variables themselves are presented in a parametric form that is difficult to penetrate. The purpose of this article is to present several perspectives where these new solutions can be interpreted as extensions of considerations that are already well-known in the literature. In this manner we study the extension of established models to include the interplay between angular momentum and intrinsic spin. 

The most basic strategy for making a connection between 4D black holes and microscopic considerations is to determine an effective string description where the black hole entropy is interpreted in terms of a dilute gas in one spatial dimension. We exhibit a limit where this description applies and the internal spin appears non-trivially. A specific advance is that we invert the parametric formulae for asymptotic charges such that the black hole entropy and the black hole mass are expressed in terms of asymptotic charges in this limit. 

For the purpose of the effective string description the energy formula we find can be recast as two conformal weights of the dual conformal field theory (CFT). These conformal weights are intricate functions of the black hole charges so it is non-trivial that they give the correct black hole entropy. The effective conformal weights were in fact already predicted using the enhanced symmetry inherent in the attractor mechanism for black strings \refs{\LarsenDH,\LarsenQR}. The entropy formula found here by explicit computation has the anticipated form but it depends implicitly on a KK-monopole charge $P^0$ that is too small to appear explicitly. 

A more ambitious strategy for making a connection between 4D black holes and microscopic considerations is inspired by the remarkable separability of scalar field equations noticed in the background of rotating ``four charge" black holes \CveticXV\ (and now 
extended to the ``fifth" charge \refs{\ChowTIA,\ChowCCA}). It reads off general conformal weights that apply beyond the dilute gas limit \CveticVP, at least for some purposes. One aspect of this idea was developed as ``hidden conformal symmetry'' \CastroFD. Another implementation introduces a ``subtracted geometry" that modifies asymptotic boundary conditions far from the black hole such that simplifications in the near horizon region become exact \refs{\CveticDN,\CveticHP}. (For further work see e.g., \refs{\CveticTR,\VirmaniKW,\BaggioDB,\ChakrabortyNU,\ChakrabortyFX,\CveticCJA,\CveticINA,\CveticLFA} and references therein.) In this article we specify the subtracted geometry for the rotating solutions including internal spin and we discuss the resulting conformal weights. 

The structure underpinning both the dilute gas description and the subtraction procedure is the geometry of a BTZ black hole in AdS$_3$ (for review see \KrausWN). It is not obvious {\it a priori} that an underlying AdS$_3$ geometry remains after intrinsic spin is added. 
We construct the sources supporting the subtracting geometry by taking a scaling limit of a general solution \CveticTR. This let us exhibit an explicit lift of the subtracted geometry to a 5D geometry that takes the form of an $S^2$ fibered over a BTZ base. We show that the  intrinsic spin modifies the mass and the angular momentum of the effective BTZ black hole in the same way as the dilute gas analysis. This verification ties together the different strategies and exhibits the details of how they survive the addition of internal spin. 

The subtraction procedure associates the general geometry with an AdS$_3$-type geometry which in turn is dual to a CFT. This CFT has identifiable parameters and so serves as a tool to interpret the general black holes. The complete specification of the dual CFT requires the addition of  large irrelevant deformations \BaggioDB, yet the parameters of the subtracted geometry description are central to the physical properties of the general black holes. 

This paper is organized as follows. In section 2 we analyze the parametric formulae for conserved charges and for the entropy. We exhibit a dilute gas limit of the geometry that incorporates intrinsic spin and reduces to the standard case in the ``four charge'' limit. We present explicit formulae for the physical variables. In section 3 we discuss the $U(1)^3$ subgroup of the duality group that acts as a symmetry on the black holes. We show that it is consistent with the dilute gas scaling limit as far as the charges are concerned, yet the overall $U(1)$ takes the solutions out of the family we consider. In section 4 we consider a CFT model of the black holes in the dilute gas limit. In this model the magnetic charges appear as $U(1)$ currents in the CFT. 

In section 5 and subsequent sections we return to the general set of charges with no dilute gas limit. We derive the thermodynamics in a manner that is independent of the complicated conformal factor $\Delta_0$. We then analyze the scalar wave equation and extract the simpler conformal factor $\Delta$ that defines the subtracted geometry. In section 6 we derive the subtracted geometry independently, by considering a scaling limit on the full solution (including matter) and then rewriting in terms of general parameters. This procedure constructs the matter sources that support the subtracted geometry. In section 7 we recast the subtracted a geometry as an $S^2$ fibered over a BTZ and compute the conformal weights of the underlying CFT description. These results generalize the dilute gas limit to the setting with general charges. 

Our concluding section 8 provides a more detailed discussion of the relation between the ``dilute gas" and ``subtracted geometry" analyses of these black holes. An appendix A summarizes the Chow-Comp{\`e}re solution and appendix B displays some useful expressions in the dilute gas approximation.

\newsec{The Dilute Gas Limit}
In this section we implement the dilute gas limit on the relation between parametric charges and physical charges. This gives simplified formulae that let us express the black hole entropy and the excitation energy explicitly in terms of physical parameters. 

%%%%%%%%%%%%%%
\subsec{The Configuration Space}
The general black hole solution depends on $10$ physical parameters: the mass $M$, the angular momentum $J$, $4$ electric charges $Q_I$ (I=0,1,2,3), and $4$ magnetic charges $P_I$ (I=0,1,2,3). The black hole geometry and the black hole thermodynamics is expressed in terms of $10$ auxiliary parameters: the parametric mass $m$, the parametric angular momentum $a$, $4$ electric boosts $\delta_I$, and $4$ magnetic boosts $\gamma_I$. 

The relation between the $10$ physical parameters and the $10$ auxiliary parameters is expressed in terms of four potentials that are complicated functions of the boosts. The two ``mass-type" potentials
\eqn\aa{\eqalign{
\mu_1  & = {1\over 4} c_{2\delta 0} \left( 1 + \sum_is^2_{\gamma i} - s^2_{\gamma 0}\right)+ {1\over 4}\left[ c_{2\delta 1} \left( 1 + s^2_{\gamma 0} - s^2_{\gamma 1}+ s^2_{\gamma 2} + s^2_{\gamma 3} \right)
+{\rm cyclic}\right]\, , \cr
\mu_2 & = {1\over 2}s_{2\delta 0}\left( s_{\gamma 0}c_{\gamma 123}-c_{\gamma 0}s_{\gamma 123}\right) + 
{1\over 2}\left[ s_{2\delta 1}\left( s_{\gamma 1}c_{\gamma 023}-c_{\gamma 1}s_{\gamma 023}\right)+ {\rm cyclic}\right]~,
}}
and the two ``angular momentum type" potentials
\eqn\ab{\eqalign{
\nu_1 & = s_{\gamma 0}c_{\gamma 0} ( c_{\delta 0} s_{\delta 123} - s_{\delta 0}c_{\delta 123})
+ \left[ s_{\gamma 1}c_{\gamma 1} ( c_{\delta 1} s_{\delta 023} - s_{\delta 1}c_{\delta 023}) + {\rm cyclic}\right]\, , \cr
\nu_2 & = (c_{\delta 0123}  - s_{\delta 0123})(c_{\gamma 0123}  - s_{\gamma 0123})\cr
&- \left[ ( c_{\delta 01}s_{\delta 23} - s_{\delta 01}c_{\delta 23} )( c_{\gamma 01}s_{\gamma 23} - s_{\gamma 01}c_{\gamma 23} )+ {\rm cyclic}\right]~. 
}}
These formulae are the general relations. They are symmetric under permutation of the four boost indices $I=0,1,2,3$ although we have written them in a form where the index ``I=0'' is stressed at the expense of $i=1,2,3$.
The notation is condensed so $s_{\delta I} = \sinh \delta_I$, $c_{\delta I} = \cosh \delta_I$, and similarly for $\gamma_I$. Several numerical indices after the greek index imply a product such that e.g., $s_{\delta 12} = \sinh \delta_1\sinh \delta_2$. Numerical indices before the greek index imply multiples of the angle such that e.g., $s_{2\delta 1} = \sinh 2\delta_1$

The physical mass $M$ and the physical angular momentum $J$ are expressed in terms of the corresponding auxiliary parameters $m$, $a$ and the potentials as 
\eqn\af{\eqalign{
G_4 M &= m(\mu_1 - {\nu_1\over\nu_2}\mu_2)~,\cr
G_4 J &= am{\nu^2_ 1 + \nu^2_2\over\nu_2}~.
}}
The physical electric charges are given by
\eqn\ac{\eqalign{
{1\over 2} Q_0 & = m( {\partial\mu_1\over\partial\delta_0} - {\nu_1\over\nu_2} {\partial\mu_2\over\partial\delta_0})~, \cr
{1\over 2} Q_i & = m( {\partial\mu_1\over\partial\delta_i} - {\nu_1\over\nu_2} {\partial\mu_2\over\partial\delta_i}) ~.
}}
Explicitly, these formulae give the charge
\eqn\ad{\eqalign{
{1\over 2} Q_0 &=m \left[  {1\over 2} s_{2\delta 0}\left( 1 + \sum_i s^2_{\gamma i} - s^2_{\gamma 0}\right)- {\nu_1\over\nu_2} c_{2\delta 0} \left( s_{\gamma 0}c_{\gamma 123}-c_{\gamma 0}s_{\gamma 123}\right) \right]~,
}}
and
\eqn\ae{\eqalign{
{1\over 2} Q_1 &= m \left[  {1\over 2} s_{2\delta 1}  ( 1 + s^2_{\gamma 0}+ s^2_{\gamma 2}+ s^2_{\gamma 3} - s^2_{\gamma_1}) -  {\nu_1\over\nu_2} c_{2\delta 1} (s_{\gamma 1} c_{\gamma 023}
- c_{\gamma 1} s_{\gamma 023})\right]~,
}}
with $Q_2, Q_3$ determined by cyclic permutations of indices $1,2,3$. The physical magnetic charges are given by
\eqn\aca{\eqalign{
{1\over 2} P_0 & = -m( {\partial\nu_1\over\partial\delta_0} - {\nu_1\over\nu_2} {\partial\nu_2\over\partial\delta_0})~,\cr
{1\over 2} P_i & = -m( {\partial\nu_1\over\partial\delta_i} - {\nu_1\over\nu_2} {\partial\nu_2\over\partial\delta_i})~. 
}}
We shall not endeavour to provide the results of the differentiations in \aca. 

%%%%%%%%%%%%%%
\subsec{The General Dilute Gas Limit}
The standard dilute gas limit is usually expressed in terms of four boost angles: take $\delta_i\gg 1$ with $\delta_0$ fixed. The strict limit $\delta_i\to\infty$ must be accompanied by taking $m\to 0$ with $m e^{2\delta_i}$ fixed. We study the dilute gas limit with all the magnetic parameters 
$\gamma_I$ kept fixed just like the boost $\delta_0$. As a practical matter we will mostly keep $m$ fixed and take 
$e^{2\delta_i}$ large without taking the strict limit. 

The angular momentum type potentials \ab\ simplify in the dilute gas limit. At the leading order we have: 
\eqn\ba{\eqalign{
\nu_1 & = c_{\delta 123} e^{-\delta_0}[ s_{\gamma0}c_{\gamma0}- \sum_i s_{\gamma i}c_{\gamma i}]+ {\cal O}(e^{\delta_i})~, \cr
\nu_2 &=
 c_{\delta 123} e^{-\delta_0}[  (c_{\gamma0123}-s_{\gamma0123})
+ (s_{\gamma01}c_{\gamma23} - c_{\gamma01}s_{\gamma23})+{\rm cyclic}]+ {\cal O}(e^{\delta_i})~.
}}
These potentials are very large in the dilute gas limit, of order $\nu_{1,2}\sim e^{3\delta_i}$; but their ratio $\nu_1/\nu_2$ is a constant of order $\sim 1$ that is independent of all $\delta_I$. The charges $Q_i$ (given by \ae\ and its cyclic permutations) also simplify in the dilute gas limit. At leading order
\eqn\bb{\eqalign{
{1\over 2} Q_1 &= m c_{2\delta 1}  \left[  {1\over 2}  ( 1 + s^2_{\gamma 0}+ s^2_{\gamma 2}+ s^2_{\gamma 3} - s^2_{\gamma_1}) -  {\nu_1\over\nu_2}(s_{\gamma 1} c_{\gamma 023}
- c_{\gamma 1} s_{\gamma 023})\right] + {\cal O}(1)~.
}}
These charges are large, of order $\sim e^{2\delta_i}$. On the other hand $Q_0$ \ad\ is small, of order 1. All these scalings are the same as the standard dilute gas limit. The generalization is that presently the dual magnetic charges are incorporated as well. 

The magnetic charges $P_I$ are expressed in \aca\ as derivatives of the magnetic potentials 
$\nu_{1,2}$. As such they would appear very large, of order $\sim e^{3\delta_i}$. However, to the extent the magnetic potentials $\nu_{!,2}$ can be approximated at leading order \ba, the magnetic charges all vanish identically because $\nu_{1}$ and $\nu_{2}$ depend on $\delta_I$ through the same overall factor $c_{\delta 123}e^{-\delta_0}$. The leading approximation for the magnetic charges is therefore obtained by computing the combinations 
\eqn\bc{
{\partial\nu_1\over\partial\delta_I}
- {\nu_1\over\nu_2}{\partial\nu_2\over\partial\delta_I}~,
}
and only then take the limit of large $\delta_I$. The resulting magnetic charges are large, but only of order $e^{\delta_i}$. The ``spatial" magnetic charges are:
\eqn\bd{
-{1\over 2} P^1 = {mc^2_{\delta 23}\over 2\nu_2} \sinh(\gamma_0-\gamma_1) \cosh(\gamma_2-\gamma_3)\left( \cosh 2(\gamma_0+\gamma_1) + \cosh 2(\gamma_2+\gamma_3)\right)~,
}
and cyclic permutations. The ``temporal" magnetic charge $P^0$ is precisely such that 
\eqn\bda{
P_0 + P_1 + P_2 + P_3=0~.
}
This is surprising. It shows that the four parameters $\gamma_I$ are redundant in the dilute gas limit: they parametrize just three independent magnetic charges. We will discuss this further in the next sections. 

For perspective on the scaling of the charges we consider the
quartic invariant of the charges 
\eqn\be{
I_4 = {1\over 16} \left[ 4Q_0 Q_1 Q_2 Q_3   + 4P_0  P_1 P_2 P_3 -
\sum_{I=0}^3 (Q_I P_I)^2+2\sum_{I<J} Q_I P_I Q_J P_J\right]~.
}
Recalling that the $Q_i$ are large $\sim e^{2\delta_i}$ while $Q_0$ is of order $\sim 1$ the product of all four electric charges is $\sim e^{6\delta_i}$. The four magnetic charges are all of order $\sim e^{\delta_i}$ so their product is just $\sim e^{4\delta_i}$ and therefore negligible at the leading order. Products of the form $Q_i P_i Q_j P_j\sim e^{6\delta_i}$ 
remain at leading order but the entire dependence on $P_0$ drops out from the quartic invariant in the dilute gas limit. Since $P_0$ is paired with the small charge $Q_0$ it would have had to be very large $P_0\sim e^{3\delta_i}$ in order to contribute but in the dilute gas limit $P_0\sim e^{\delta_i}$ like the $P_i$'s. Thus $P_0$ does not appear in the quartic invariant \be\ which simplifies to
\eqn\bg{
I_4 = {1\over 16} \left[ 4Q_1 Q_2 Q_3 Q_0 -
\sum_{i=1}^3 (Q_i P_i)^2+ 2\sum_{i<j} Q_i P_i Q_j P_j \right]~.
}

\subsec{Black Hole Mass}
In the dilute gas limit the total mass of the black hole \af\ is large, of order $\sim e^{2\delta_i}$, but the BPS mass
\eqn\bm{
M_{\rm BPS} =  {1\over 4}\sqrt{ (\sum_I Q_I)^2  +  (\sum_I P_I)^2 } ~,
}
has identical asymptotic behavior. We can therefore introduce the finite excitation energy $E$ through
\eqn\bma{
E - {\textstyle{1\over 4} }Q_0 = M - M_{\rm BPS}~. 
}
The sum of magnetic charges vanishes \bda\ so we have
\eqn\bh{\eqalign{
E & =  M - {1\over 4} \sum_i Q_i \cr
&= m (\mu_1 - {1\over 2}\sum_i {\partial\mu_1\over\partial\delta_i}) - m {\nu_1\over\nu_2}(\mu_2 - {1\over 2}\sum_i {\partial\mu_2\over\partial\delta_i})\cr
& ={m\over 4} c_{2\delta 0} \left( 1 + \sum_i s^2_{\gamma i} - s^2_{\gamma 0}\right)
- {\nu_1\over\nu_2}{m\over 2}s_{2\delta 0}\left( s_{\gamma 0}c_{\gamma 123}-c_{\gamma 0}s_{\gamma 123}\right)+{\cal O}(e^{-2\delta_i})~,
}}
in units where $G_4=1$. In this formula the ratio $\nu_1/\nu_2$ can be computed using the leading terms given in \ba. Although 
$M$ and $Q_i$ all have large exponentials (of order $e^{2\delta_i}$) the leading contributions indeed cancel so that the excitation energy $E$ is finite. 

We can simplify the chiral energy as 
\eqn\bi{\eqalign{
 E- {\textstyle{1\over 4}}Q_0    = {me^{-2\delta_0}\over4 [\nu_2]} &\cosh(\gamma_0+\gamma_1-\gamma_2-\gamma_3)\cosh(\gamma_0-\gamma_1+\gamma_2-\gamma_3)\cr
 &
 \times \cosh(\gamma_0-\gamma_1-\gamma_2+\gamma_3)\, .}}
The notation $[\nu_2] = \nu_2 / (e^{-\delta_0}c_{\delta 123})$ isolates the $\gamma$-dependent factors in the square bracket of \ba. 
The analogous expressions for the large charges are
\eqn\biw{\eqalign{{\textstyle {1\over 2}} Q_1   & = {mc_{2\delta 1}\over 2 [\nu_2]} \cosh(\gamma_0+\gamma_1+\gamma_2+\gamma_3)\cosh(\gamma_0-\gamma_1+\gamma_2-\gamma_3)\cosh(\gamma_0-\gamma_1-\gamma_2+\gamma_3)~,
}}
and cyclic permutations. Our conventions are such that the BPS limit corresponds to all $Q_I>0$. According to \biw\ this is only possible if the $\gamma_I$ are such that $\nu_2>0$. The right hand side of the energy formula \bi\ is therefore manifestly positive and so in agreement with the BPS bound $E\geq {1\over 4}Q_0$. The bound is saturated in the  BPS limit $\delta_0\to\infty$. 

\subsec{Black Hole Entropy}
We will discuss the geometry underlying black hole thermodynamics in section 4. 
For now we simply quote the black hole entropy (in units where $G_4=1$)
\eqn\bia{\eqalign{
S= 2\pi\sqrt{ F - J^2}  + 2\pi\sqrt{F + I_4}~, 
}}
where 
\eqn\bj{
F\equiv m^4  {(\nu_1^2 + \nu_2^2)^3\over\nu_2^4}~. 
}
The dependence of the quantity $F$ on physical charges is complicated due to the dependence \ab\ of $\nu_{1,2}$ on the parametric charges $\delta_I, \gamma_I$ and also the relation between parametric charges and physical charges. 

In order to simplify $F$ in the dilute gas limit we first compute 
\eqn\bm{\eqalign{
\nu_1^2 + \nu_2^2 
&= c^2_{\delta 123} e^{-2\delta_0}\cosh(\gamma_0+\gamma_1-\gamma_2-\gamma_3)\cosh(\gamma_0+\gamma_1-\gamma_2-\gamma_3) \cr & ~~~~~~~\times\cosh(\gamma_0-\gamma_1-\gamma_2+\gamma_3)
\cosh(\gamma_0+\gamma_1+\gamma_2+\gamma_3)~.
}}
Comparing with the chiral excitation energy \bi\ and the electric charges in the form \bia\ we then find
\eqn\bn{\eqalign{
F &= 
m^4 c^2_{\delta 123}e^{-6\delta_0}{1\over\nu_2^4}\left[ \cosh(\gamma_0+\gamma_1-\gamma_2-\gamma_3)\cosh(\gamma_0+\gamma_1-\gamma_2-\gamma_3)\right. \cr & \left. ~~~~~~~\times\cosh(\gamma_0-\gamma_1-\gamma_2+\gamma_3)
\cosh(\gamma_0+\gamma_1+\gamma_2+\gamma_3)\right]^3\cr
& = {1\over 2} Q_1 Q_2 Q_3 (E- {1\over 4} Q_0)  ~,
}}
in the dilute gas limit. We can therefore recast the black hole entropy \bia\  as
\eqn\bo{\eqalign{
{S\over 2\pi}&= \sqrt{
{Q_1Q_2Q_3\over 2}   (E-
{Q_0\over 4}) - J^2} \cr
&+ 
\sqrt{
 {Q_1Q_2Q_3\over 4}(E + 
 {Q_0\over 4})  - 
 {\textstyle{1\over 16}}\sum_{i=1}^3 (Q_i P_i)^2+ {{\textstyle{1\over 8}}}\sum_{i<j} Q_i P_i Q_j P_j }\, . }
}
Recall that the excitation energy $E$ defined in \bma\ is essentially the physical mass. The formula \bo\ therefore expresses the entropy in terms of physical mass, physical angular momentum, and physical charges. This is remarkable because the original entropy formula \bia\ was cast in terms the parametric mass and charges which are related quite non-trivially to physical parameters. 

The dependence on magnetic charges in the final result \bo\ is such that in the BPS limit $E\to {1\over 4}Q_0$ we reproduce the BPS result known from the attractor mechanism and from explicit solutions \refs{\BehrndtJN,\ShmakovaNZ,\MaldacenaDE}. From this perspective the dependence of the entropy formula on magnetic charges appears unremarkable and nearly trivial. The novel feature is the implicit dependence on the magnetic charge $P_0$ given by \bda. 

\newsec{Global Symmetry}
In this section we discuss the $U(1)^3$ symmetry preserved by the dilute gas limit. The symmetry acts nontrivially on the dilute gas solutions and permits consideration of an enlarged family of configurations. 

\subsec{The $U(1)^3$ symmetry} 
The $8$ charges of the STU model transform in the $(2,2,2)$ of the $[SL(2)]^3$ duality group. The duality transformations generally act on the three complex scalar fields parametrizing the $[SL(2)/U(1)]^3$ coset but the $U(1)^3$ ``denominator" subgroup leaves the scalars invariant. This $U(1)^3$ acts as a nontrivial symmetry of the theory. Because of this $3$ parameter symmetry the most general asymptotically flat solution is parametrized up to duality by just $8-3=5$ charge parameters. Equivalently, there is in principle a $3$ parameter redundancy in the $8$ parameters $(\delta_I, \gamma_I)$ employed by Chow and Comp{\`e}re. However, the action of these symmetries is notoriously nontrivial so it is worthwhile to consider the details.

The three angles $\theta_i$ that parametrize the $U(1)^3$ symmetry generically violate the hierarchy of charges assumed in the dilute gas limit but parametrically small angles $\theta_i\sim e^{-\delta_i}$ act nontrivially within this configuration space. They transform the charges as
\eqn\bq{\eqalign{
P'_0 & = P_0 -  (\theta_1 Q_1 + \theta_2 Q_2+ \theta_3 Q_3)~,\cr
Q'_0 & = Q_0 + (\theta_1 P^1 + \theta_2 P^2+\theta_3 P^3) - (\theta_1 \theta_2 Q_3+ {\rm cyclic})~, \cr
P'_1 & = P_1- (\theta_2 Q_3 + \theta_3 Q_2)~, ~({\rm and~cyclic~permutations})~.
}}
We keep terms of the same order as the charges involved: $P_I\sim e^{\delta_i}$ and $Q_0\sim 1$. The large charges $Q_i\sim e^{2\delta_i}$ are invariant at leading order so we shall mostly consider them invariant. However, it is worth recording that at order $\sim 1$ these charges transform in a complicated way:
\eqn\bra{
Q'_1 = (1 - {1\over 2} (\theta_1^2 + \theta_2^2+ \theta_3^2)  )Q_1 - \theta_1 (\theta_2 Q_3+ \theta_3 Q_2 ) +  (\theta_2 P_3 + \theta_3 P_2) + \theta_1 P_0 ~,
}
with $Q'_2, Q'_3$ given by cyclic symmetry. 

The $U(1)^3$ transformations \bq\ act on magnetic charges so 
\eqn\bqd{
(P_0 + P_1 + P_2 + P_3)' = (P_0 + P_1 + P_2 + P_3) - (\theta_1 + \theta_2 + \theta_3) (Q_1 + Q_2 + Q_3) ~. 
}
Thus the sum rule \bda\ for magnetic charges $\sum_I P_I=0$ found by explicit computations does not respect the $U(1)^3$ symmetry. In particular, the action of the diagonal $U(1)$ symmetry cannot be realized as a transformation of the $\gamma_I$'s; it involves the $\delta_I$'s as well. However, the entropy 
formula \bo\ is correct in any $U(1)^3$ frame as long as the excitation energy $E$ is measured with respect to the full BPS ground state mass, as in \bma. In contrast, the evaluation \bh\ relies on the sum rule \bda\ and so it is not $U(1)^3$ invariant. 

To verify this explicitly note that the black hole mass and the BPS mass are both invariant under duality so the excitation energy $E-{1\over 4}Q_0=M-M_{\rm BPS}$ is invariant as well. The entropy formula \bo\ is also $U(1)^3$ invariant because 
\eqn\bqa{
I_0 = Q_0  - {1\over 4Q_1 Q_2 Q_3} \sum_{i=1}^3 (Q_i P_i)^2+ {1\over 2Q_1 Q_2 Q_3}\sum_{i<j} Q_i P_i Q_j P_j ~,  
}
is invariant. We can understand the explicit dependence of the entropy on the magnetic charges from this property. Starting from the dilute gas solution with general $P_i$ we can transform to $P'_i=0$ by selecting $U(1)^3$ angles
\eqn\bqb{
\theta_1 = {-P_1 Q_1 + P_2 Q_2 + P_3 Q_3\over 2 Q_2 Q_3}~,~({\rm and~cyclic~permutations})~.
}
The $I_0$ invariant \bqa\ shows that the dependence of the entropy formula \bo\ on the magnetic charges is absorbed in the momentum charge $Q_0$. Conversely, the $U(1)^3$ symmetry determines the full entropy formula \bo\ from the special case without magnetic charges. 

\subsec{The Ground State Energy} 
The $U(1)^3$ independence of the explicit entropy formula hides an implicit dependence. Indeed, the $U(1)^3$ transformation to $P'_i=0$ does not leave the full system invariant since it also changes $P_0$ to 
\eqn\bqc{\eqalign{
P_0' & = P_0 + \left( P_1 {Q_1^2 - Q_2^2-Q_3^2\over 2Q_2 Q_3} + {\rm cyclic} \right) \cr
& = P_1 {Q_1^2 - (Q_2+Q_3)^2\over 2Q_2 Q_3} + {\rm cyclic}  ~. 
}}
The right hand side of the first line records an interesting $U(1)^3$ invariant. The second equation employed $P_0=-\sum_i P_i$. 

In the limit we consider the natural ground state is the background with the large charges $Q_i$ since all dependence on magnetic charges constitutes finite energy excitations of that theory. From this perspective the ground state energy energy is larger than the excitation energy $E$ above the BPS mass (defined in \bma)
\eqn\btc{
E_{\rm gs} = M - {1\over 4} \sum_i Q_i = E + {(\sum_I P_I)^2\over 8\sum Q_I}~.
}
This shift is inconsequential for the explicit solutions where $\sum_I P_I=0$ but it is significant in a general $U(1)^3$ frame. The transformation \bqb\ that removes the magnetic charges $P'_i=0$ indeed shifts $\sum_I P_I =0$ to $\sum_I P'_I = P'_0\neq 0$. This gives a ground state energy
\eqn\bqd{
E_{\rm gs} - E= {1\over 8(Q_1 + Q_2 + Q_3) } P_0^{\prime 2} = {Q_1 + Q_2 + Q_3 \over 8}\left[ P_1 {Q_1 - Q_2-Q_3\over 2Q_2 Q_3} + {\rm cyclic} \right]^2~, 
}
where we used \bqc. The same result can be derived (less easily) from the transformation \bra\ of the large electric charges. 

The energy \bqd\ assigned to magnetic charges by this procedure is somewhat nontrivial. For example, 
\eqn\bqe{
E_{\rm gs} - E = {27\over 32}{P^2\over Q}~,
}
for diagonal charges. For inspiration, we recall the corresponding energy due to the addition of a magnetic brane to its dual electric background brane \refs{\RasheedZV,\LarsenPP,\LarsenPU}
\eqn\bqf{
M - M_{\rm BPS} = {1\over 4} (Q^{2/3} + P^{2/3} )^{3/2} - {1\over 4} Q = {3\over 8}P^{2\over 3}Q^{1\over 3}  ~.
}
In the present context we have three types of electric background branes and we add their three dual magnetic charges. As a result the energy \bqe\ has acquired additional factors of $3$. 

\newsec{A Microscopic Model}
In this section we interpret our results in the framework of the MSW $(4,0)$ CFT \MaldacenaDE. We track the conventions between physical charges and quantized charges; we discuss the attractor mechanism for the effective string description; and we describe the magnetic charges in terms of $U(1)$ currents in the CFT. 

\subsec{Quantized Charges and Moduli}
The relation between physical charges and quantized microscopic charges can be established without loss of generality in the context of an elementary non-rotating extremal black hole with just four charges. 

All charges $Q_I, P_I$ in this work are lengths and normalized as the residue of the harmonic function $H_I = 1 + {Q_I\over r}$. 
Such a charge is 
related to the mass of the corresponding isolated brane as $M={1\over 4G_4}Q$. The factor relating the mass and the quantized charge is just the tension and volume. For example, $q$ M5-branes wrapping a four-cycle (with volume $V$) times a circle (with length $2\pi R$) have mass
\eqn\bkx{
M_{\rm M5} = {R V\over (2\pi)^4 l_P^6}\cdot q~.
}
These conversion factors combine with the 4D Newton coupling $G_4 = {l_P^9(2\pi)^6\over 8R{\rm Vol}_6}$ such that the product of four physical charges compatible with SUSY convert to quantized charges as 
\eqn\bky{
Q_1 Q_2 Q_3 Q_4 = 4 G^2_4 q_1 q_2 q_3 q_4~. 
}

Supergravity formulae such as \bh\ and \bm\ for the mass 
take the gravitational coupling constant $G_4=1$. For a single charge at the self-dual point in moduli space we can thus use the rule $Q \to \sqrt{2}q$ to translate from physical to quantized charges. Dimensionless moduli can be restored if needed and then interpreted as the physical moduli expressed relative to the self-dual point. 

The analogous considerations for the mass $M={1\over 4G_4}Q$ 
amounts to the substitution $E\to \epsilon/2\sqrt{2}$. This normalizes the excitation energy 
in units of the natural scale $G_4/Q_1 Q_2 Q_3$. Equivalently, the microscopic energy $\epsilon$ is such that $\epsilon=q_0$ in the extreme limit where $q_0$ is the canonical momentum quantum number defined through
${1\over 4G_4}Q_0 = {q_0\over R}$.  
This is natural in the effective CFT where $q_0$ is interpreted as the momentum quantum number. 

The transformation to microscopic variables takes
\eqn\bma{
{1\over 2} Q_1 Q_2 Q_3 (E\pm  {1\over 4}Q_0) = {1\over 2} q_1 q_2 q_3 (\epsilon\pm  q_0) ~,
}
so the entropy formula \bo\ becomes
\eqn\ga{
S = 2\pi \left[ \sqrt{  q_1 q_2 q_3 h^{\rm irr} _R } + \sqrt{ q_1 q_2 q_3h^{\rm irr}_L}\right]~,
}
where the irreducible conformal weights are
\eqn\gb{\eqalign{
h^{\rm irr}_R &= {1\over 2} (\epsilon-q_0)  -{1\over q_1 q_2 q_3 }J^2 ~,\cr
h^{\rm irr}_L &= {1\over 2} (\epsilon+q_0)  - {1\over 4q_1 q_2 q_3 }[  \sum_i (q_i p_i )^2 - 2\sum_{i<j} (q_i p_i)(q_j p_j) ] ~.
}}

\subsec{The Attractor String}
The microscopic description is the theory of the three large electric charges $Q_1, Q_2, Q_3$. The excitation energy, the small electric charge $Q_0$ and the magnetic charges are carried by excitations in that theory. The simplest duality frame is where the three large charges are M5-branes each wrapped on a four cycles with size of order Planck length times a large circle with size $2\pi R$. Then the mass of the 
background is the BPS mass
\eqn\gca{
M = {R\over l_P^2} [ v_1 q_1 + v_2 q_2 + v_3 q_3]~,
}
where volumes of four cycles are in Planck units $v_1=V_1/(2\pi l_P)^4$. The background mass is large because $R/l_P$ is large, 

Variations in the relative volumes of the four-cycles 
generally lowers the large background mass and so correspond to instabilities of the theory rather than parameters. We must therefore 
fix these ratios at their attractor points where 
\eqn\gcb{
{v_1\over(v_1 v_2 v_3)^{1\over 3}} = \left({q_2 q_3\over q^2_1}\right)^{2/3}~,~~({\rm and~cyclic~permutations})~. 
}
For these values of moduli the physical background charges are identical $Q_1=Q_2=Q_3$. On the other hand, the 
physical magnetic charges $P_i$ correspond to $M2$ branes charge of the form
\eqn\gcc{
P_i = {l^2_P \over 2R} {1\over v_i} p_i ~, 
}
where $v_i$ is the volume of the four-cycle dual to the two-cycle wrapped by the $M2$. The attractor mechanism modifies the rule $P_i \to\sqrt{2}p^i$ with moduli at the self-dual point to $P_i \to \sqrt{2} p^i q^i/(q_1 q_2 q_3)^{1\over 3}$. The attractor mechanism for the effective string is discussed in \LarsenDH. 

It is a consequence of the attractor mechanism that the theory depends on only the product of background charges rather than each charge by itself. For example, the central charge is 
\eqn\gcf{
c = 6q_1 q_2 q_3= 6k~.
}
This is indeed a feature of the MSW theory \MaldacenaDE.

The novel magnetic charge $P_0$ corresponds to KK monopoles and requires special considerations. After restoring units it is 
\eqn\gcd{
P_0 = {R\over 2} p_0. 
}
Since $R/l_P\gg 1$ this is much larger than the other magnetic charges \gcc\ unless the quantum number $p_0$ is taken parametrically small $p_0\sim l^2_P/R^2$. That is strange because the integer $p_0$ cannot become arbitrarily small. Equivalently, in the strict scaling limit where the circle of radius $R$ decompactifies the KK monopole does not exist. The standard option is to take $P_0=0$ from the outset but this is not possible presently since the solution specifies the sum rule \bda\ $\sum_I P_I=0$ corresponding to
\eqn\gce{
{R^2\over l_p^2} p_0 + {\vec{q} \cdot \vec{p} \over (q_1 q_2 q_3)^{1/3}} =0~. 
}
The coupling to $P^0$ is an interesting feature of the intrinsic spin $\vec{q}\cdot\vec{p}$. It has not been noticed in previous studies. For example, the attractor values of scalars in the BPS black hole give values 
\eqn\gcf{
\chi^i={P^i\over 2(Q_1 Q_2 Q_3)^{1/3}}~,
}
 and $e^{-\varphi}=\sqrt{I_4}/2(Q_1 Q_2 Q_3)^{1/3}$ that are independent of $P^0$.  

\subsec{Irreducible Conformal weight}
The irreducible conformal weight refers to the chiral energy available for excitations in such a manner that Cardy's asymptotic formula for the entropy applies. 
A standard construction \BreckenridgeIS\ realizes a global $U(1)$ charge as a chiral CFT current
\eqn\gc{
J = \sqrt{2k} \partial\varphi~. 
}
with a canonical normalization referred to as level $k$. The states in the CFT can then be taken of the form 
\eqn\gd{
V  = V_{\rm irr} V_{U(1)}~,
}
where operators with integral $U(1)$ quantum number $F$ 
\eqn\gd{
V^{U(1)} = e^{{i \over\sqrt{2k}}F\varphi}~,
}
have conformal weight 
\eqn\gf{
h_{U(1)}={1\over 4k} F^2~.
}
Thus the weight available for general excitations is reduced from $h$ to the irreducible weight $h_{\rm irr} = h - h_{U(1)}$. 

In the case of angular momentum $J$ the current is simply a $U(1)$ component of the $SU(2)$ R-symmetry that resides in the holomorphic (R) sector. The dependence of the entropy on the angular momentum enters entirely through the holomorphic sector and it is fully accounted for by this construction. Of course the conventional normalization for angular momentum allows for both half-integral and integral spin so 
the appropriate identification is $J = {1\over 2}F$. 

The magnetic charges are quantum numbers of the $U(1)^3$ symmetry so we can model them similarly. It is useful to rewrite the left moving irreducible weight \gb\ as
\eqn\gfa{ 
h^{\rm irr}_L = {1\over 2} (\epsilon+q_0)+
{1\over 12k}(q_1 p_1 + q_2 p_2+q_3 p_3)^2]   - {1\over 4k }(q_1 p_1 - q_2 p_2 )^2  -{1\over 12k}(q_1 p_1 + q_2 p_2-2q_3 p_3)^2~.
}
The last two terms on the right hand side can be understood as two canonically normalized anti-holomorphic (L) currents currents corresponding to the relative $U(1)$ symmetries. Indeed, assuming that $q_{1,2}$ are relatively prime, we can choose a basis where the charges are $F_1 = p_1 q_1 - p_2 q_2$ and $F_2 = {1\over\sqrt{3}}(p_1 q_1 + p_2 q_2 - 2p_3 q_3)$ so the conformal weight \gf\ for these two $U(1)$ currents becomes
\eqn\ggb{
h_1 + h_2 = {1\over 4k} ( p_1q_1  - p_2q_2)^2 + {1\over 12k} ( p_1q_1 + p_2q_2 - 2p_3q_3)^3 = {1\over 3k} [ \sum_i (p_iq_i)^2  - \sum_{i<j} (p_iq_i)(p_jq_j)]~.
}
Further, after embedding of the STU-model into theories with more symmetry (such as $N=8$ SUGRA) these two $U(1)$ currents are related by global symmetries to other and simpler currents \refs{\LarsenDH,\LarsenQR}. 
These other currents have canonical level so these symmetries confirm the assignment we find here. 

\subsec{Intrinsic Spin}
The diagonal $U(1)$ current $\vec{q}\cdot\vec{p}$ is the ``fifth charge" that we refer to as internal spin. This $U(1)$ current requires special considerations. It enters the entropy through the last term in the irreducible weight \gfa. This contribution has the opposite sign from the currents that correspond to the relative $U(1)$ symmetries. Further, it is only half as large or, equivalently, it has level that is twice the canonical level.

Recall that the excitation energy $\epsilon$ is the microscopic version of the excitation energy $E$ defined in \bma\ as the total mass relative to the BPS mass. Presently we consider the states diagonal $U(1)$ charge as excitations of a ground state with no magnetic charge. It is therefore appropriate to employ the ground state energy \btc\ that is larger than the excitation energy by the contribution of the magnetic charges to the BPS mass. Guided by the attractor mechanism in the extremal limit we assume that the theory is independent of $P^0$ and so take $P^0=0$ without loss of generality. After transcription to microscopic terminology the ground state energy then becomes 
\eqn\boq{
\epsilon_{\rm gs} = \epsilon   +  {(\vec{q} \cdot \vec{p})^2\over 6k} ~.
}
The entire contribution of the diagonal $U(1)$ to the left moving conformal weight \gfa\ is taken into account by this shift. Conversely there will be a negative contribution to the irreducible weight $h_R^{\rm irr}$. This we can model by an anti-holomorphic $U(1)$ current with twice 
the canonical level. 

A more complete microscopic model would derive the $p_0$ independence without resort to a space-time argument. As a matter of principle all of $p_0$, $\vec{q}\cdot\vec{p}$, and the asymptotic angular momentum $J$ are independent observables in the asymptotic region. It is worth investigating the interplay between these variables further.

%%%%%%%%%%%%%%%%%%%%%%%%%%%
\newsec{Subtracted Geometry}
The black hole geometry encodes thermodynamics. We present the black hole thermodynamics in a form where the independence on a certain warp factor is manifest. We then determine the subtracted warp factor that yields  manifest conformal symmetry. 

\subsec{The Reduced Angular Potential}
For easy comparison with our previous work \CveticDN\ we recast the Chow-Comp\`ere metric \refs{\ChowTIA,\ChowCCA} as
\eqn\fa{\eqalign{
ds^2  & = -\Delta_0^{-{1\over 2}} G ( dt + {\cal A})^2
+ \Delta_0^{1\over 2}\left( {dr^2\over X} + d\theta^2 + {X\over G}\sin^2\theta d\phi^2\right)~. 
}}
This amounts to a simple change of notation from \refs{\ChowTIA,\ChowCCA}:
\eqn\fb{\eqalign{
X(r) & = R(r)~,\cr
\Delta_0(r,\theta) & = W^2(r,\theta)~,\cr
G(r,\theta) & = X(r) - a^2 \sin^2\theta = R(r) - U(\theta)~,\cr
{\cal A}(r,\theta) &= \omega_3 (r,\theta)~.
}}
The functions
\eqn\fg{\eqalign{
R(r) &= X(r) = r^2 - 2mr + a^2 - {m^2\nu^2_1\over\nu^2_2}~,\cr
U(\theta) &= a^2\sin^2\theta~.
}}
Here and in the following we preserve some redundancy in an attempt to facilitate comparison with either notation. 

In the metric \fa\ rotation is encoded in the one-form ${\cal A}(r,\theta)$ on the base. The $\theta$-dependence of this one-form is such that 
\eqn\fc{
{\cal A}_{\rm red}(r) = a {R(r)-U(\theta)\over U(\theta)} {\cal A}_\phi =  L(r)~,
}
depends on $r$ alone. 
Metrics of the general form \fa\ with the functions $\Delta_0(r,\theta),$ $X(r),$ ${\cal A}_{\rm red}(r)$ unspecified were analyzed in \CveticDN. It was found that the black hole entropy can be expressed as 
\eqn\fd{
S = {\pi\over G_4} {\cal A}_{\rm red, hor} = {\pi\over G_4} L_{\rm hor}~.
}
The remarkable feature of this formula is that it only depends on the reduced angular potential ${\cal A}_{\rm red}$ ($=L(r)$ in Chow-Comp\`ere notation). In particular, the entropy does not depend on the warp factor $\Delta_0$. From this perspective the non-rotating case is a somewhat singular limit where a zero in the true angular potential ${\cal A}_\phi$ is cancelled by the overall factor taken out in \fc.

In the Chow-Comp\`ere solution 
\eqn\fe{
L(r) = 2m {\nu^2_1 + \nu^2_2\over\nu^2_2} \left( \nu_2 r + 2m D\right)~,
}
where
\eqn\ff{
D= c_{\delta 0123} s_{\gamma 0123} + s_{\delta 0123} c_{\gamma 0123} + 
( c_{\delta 01} s_{\delta 23}  s_{\gamma 01} c_{\gamma 23}  + 5~{\rm permutations} ) ~.
}
The horizons of the black holes are situated at the zeros of $X(r)$ \fg\ 
\eqn\fh{
r_\pm = m \left( 1 \pm {1\over\nu_2}\sqrt{\nu^2_1 + \nu_2^2 - a^2\nu_2^2/m^2}\right)~.
}
The formula \fd\ for the inner and outer horizon entropies \CveticUW\ therefore work out to 
\eqn\fia{\eqalign{
S_\pm & = \pi L(r_\pm) = 2\pi (\nu^2_ 1 + \nu^2_2)\left( {m\over\nu_2} r_\pm + 2D {m^2\over\nu_2^2}\right) \cr
& = 2\pi (\nu^2_ 1 + \nu^2_2){m^2\over\nu_2^2} \left( (\nu_2+ 2D) \pm  \sqrt{\nu^2_1 + \nu_2^2 - a^2\nu_2^2/m^2} \right) \cr 
& = 2\pi \left((\nu^2_ 1 + \nu^2_2){m^2\over\nu_2^2}  (\nu_2+ 2D) \pm  \sqrt{{m^4(\nu^2_1 + \nu_2^2)^3\over\nu_2^4} - a^2m^2(\nu^2_ 1 + \nu^2_2)^2/\nu_2^2 } \right)\cr
& = 2\pi\left( \sqrt{F + I_4} \pm \sqrt{ F - J^2}\right)~,
}}
in units where $G_4=1$. The notation $F$ was given in \bj\ and 
\eqn\fl{\eqalign{
I_4 & = m^4 {(\nu^2_ 1 + \nu^2_2)^2\over\nu_2^4}  \left( (\nu_2+ 2D)^2 - (\nu_1^2 + \nu_2^2)\right) 
~.
}}
This formula for $I_4$ is identical to \be\ after insertion of expressions for $\nu_{1,2}, D$ that introduce parametric charges $\delta_I, \gamma_I$ which are subsequently eliminated in favor of physical charges. The entropy $S_+$ computed here justifies \bia. 

\subsec{Subtracted Geometry}
The black hole entropy \fd\ and all other thermodynamic variables are independent of the warp factor 
$\Delta_0$. We therefore interpret the warp factor as a feature of external environment that is irrelevant for the internal structure of the black hole. The subtracted geometry is an auxiliary geometry where the warp factor has been modified $\Delta_0\to\Delta$ with 
$\Delta$ determined such that the near horizon structure simplifies. 

Separability of the scalar wave equation is an aesthetically pleasing property of the full geometry that we would like to preserve. It amounts to the requirement that the effective potential
\eqn\ha{
{\Delta_0 - {\cal A}_{\rm red}^2\over G(r,\theta)} = {W^2(r,\theta) -L^2(r)\over R(r)-U(\theta)} = (R-U) + 2(2Mr + V) = X(r) + 4Mr + 2V(r) - a^2\sin^2\theta~,
}
is a function of $r$ plus a function of $\theta$. This is indeed the case in the Chow-Comp\`ere solution. The function $V(r)$ defined through \ha\ is a quartic polynomial in $r$ with detailed form that we will not need. 
The subtracted geometry modifies the conformal factor $\Delta_0\to\Delta$ such that separability is preserved but $\Delta$ increases just linearly in $r$. 

The reduced angular potential ${\cal A}_{\rm red}(r)=L(r)$ is given in \fe\ as a linear function $L=L_1 r + L_2$ where
\eqn\haa{\eqalign{
L_1 & = 2m{\nu_1^2+\nu^2_2\over\nu_2}~,\cr
L_2 & = 4m^2{\nu_1^2+\nu^2_2\over\nu^2_2} D~. 
}}
There is a unique way to decompose its square as a term proportional to $G(r,\theta)=R(r)-U(\theta)$ and a remainder that is a linear function in $r$:
\eqn\hb{\eqalign{
{\cal A}_{\rm red}^2 = L^2 &= L_1^2 (R + 2mr - a^2 + m^2 {\nu_1^2\over\nu^2_2}) + 2L_1 L_2 r + L_2^2 \cr
&= L_1^2 (R-U) 
+  2L_1 (mL_1+ L_2 )r  + L_1^2 (U - a^2 + m^2 {\nu_1^2\over\nu^2_2} )  + L_2^2 ~.
}}
We thus take 
\eqn\hc{
\Delta = W_{\rm subtracted}^2 = 2L_1 (mL_1+ L_2 )r  + L_1^2  (U - a^2 + m^2 {\nu_1^2\over\nu^2_2} )+ L_2^2 ~.
}
The effective potential \ha\ 
\eqn\hd{
 {\Delta -{\cal A}_{\rm red}^2 \over R-U} = - L^2_1 = - {4m^2\over\nu_2^2} (\nu_1^2 + \nu_2^2)^2~,
}
for the subtracted geometry is a constant in space which is much stronger than the minimum needed for seperability. 

The subtracted geometry preserves the thermodynamics of the full solution. In that sense the black hole is not essentially different from the original and very general solution. On the other hand the subtracted solution can alternatively be interpreted as a configuration in the dilute gas limit. This is significant because it provides a path towards a microscopic description of the general black holes. 

%%%%%%%%%%%%%%%%%%%%%%%%%%%%%%%%%%
\newsec{Subtracted Geometry as a  Scaling Limit}

In this section we derive the subtracted geometry as a scaling limit of a dilute gas solution. This is an application of the strategy explained in \CveticTR\ to this more general setting. The construction identifies the matter sources that support the solution: i.e. 
the three axio-dilaton fields and the four  gauge potentials of the STU-model.

\subsec{The Scaling Limit}
The subtracted geometry is defined for the most general parameters $(m,a)$ and $(\delta_I,\gamma_I)$. However, it is an attractor in the space of solutions so it only depends on particular combinations of these parameters. For example, we may parametrize the possible subtracted geometries by ($m,\, a$) and  ($\nu_1,\  \nu_2,\   D$). A restricted set of solutions will therefore be sufficient to map out the most general subtracted geometries which in turn will correspond to the most general values of physical parameters. 
 
We parametrize a convenient and sufficient family with ``tilde'' quantities as follows. We take the charges of the three gauge potentials equal with ${\tilde \delta_i}\equiv{\tilde\delta}$ and ${\tilde \gamma_i}\equiv {\tilde\gamma}$ ($i=1,2,3$). We take 
$\tilde{\gamma}_0=0$ but we maintain general $\tilde{\delta}_0$. In addition to the three charge parameters $({\tilde\delta}, {\tilde\gamma}, \tilde{\delta}_0)$ the family is parametrized by ($\tilde{m},\,\tilde{a}$). We next introduce a scaling parameter $\epsilon$ through the change of variables 
\eqn\scaling{\eqalign{ 
&{\tilde r}= r  \epsilon,   \quad{\tilde t}= t{\epsilon^{-1}},  \quad  {\tilde m}=  m \epsilon\, , \quad  {\tilde a}=  a \epsilon\, , \cr
& \sinh^3 {\tilde \delta}  = A{\epsilon^{-2}},   \quad e^{2{\tilde \delta}_0}=B~, ~\sinh{\tilde \gamma}=C\, .
}}
For any value of $\epsilon$ the representative family of solutions is thus parametrized by $(m,a)$ and either $({\tilde\delta}, {\tilde\gamma}, \tilde{\delta}_0)$ or 
$(A,B,C)$. 

We consider the scaling limit $\epsilon \to 0$ taken with fixed coordinates $(r,t)$ and fixed parameters $(m,a)$, $(A,B,C)$. The scaling limit has been designed so charges scale to the dilute gas regime. Thus the representative configurations can be analyzed in terms of a CFT. 
Further, the coordinates are being rescaled precisely so that the full geometry reduces to the subtracted geometry. Since the configurations we focus on are in fact solutions the matter supporting them is specified explicitly by taking the appropriate special case of the matter in the general Chow-Comp\` ere black holes.

Now, we associate a subtracted geometry with a completely general black hole. We want to find the map between the general parameters 
($\nu_1,\  \nu_2,\   D$) and the $(A,B,C)$ or $({\tilde\delta}, {\tilde\gamma}, \tilde{\delta}_0)$ of the representative family. The most convenient way to do so is
to consider the simplified formulae that express 
$({\tilde \nu}_1,\  {\tilde \nu}_2, \  {\tilde D})$ in terms of $({\tilde\delta}, {\tilde\gamma}, \tilde{\delta}_0)$ in the scaling limit
\eqn\scIII{\eqalign{{\tilde \nu}_1&\sim -3\cosh{\tilde \gamma}\sinh{\tilde \gamma}e^{-{\tilde \delta}_0}\cosh ^3{\tilde \delta}~,  \cr
{\tilde \nu}_2&\sim (\cosh^2{\tilde \gamma}-3\sinh^2{\tilde \gamma})e^{-{\tilde \delta}_0}\cosh^3{\tilde \delta}~, \cr   
{\tilde D}& \sim (\cosh^2{\tilde \gamma}\sinh{\tilde \delta_0}+3\sinh^2{\tilde \gamma\cosh{\tilde \delta_0})\cosh{\tilde \gamma}}\cosh^3{\tilde \delta}\,. } }
These particular combinations of parameters in the representative family of subtracted geometries can be converted to the 
general parameters through
\eqn\scII{{\tilde \nu}_1=\nu_1\epsilon^{-2}\, , \quad {\tilde \nu}_2=\nu_2\epsilon^{-2}\, , \quad {\tilde D}=D\epsilon^{-2}\, .} 
After all, the general subtracted geometry only depends on physical parameters through ($\nu_1,\  \nu_2,\   D$) so it is 
sufficient to find the required identifications in a special case. 

An equivalent procedure that is more in line with the logic presented above trades the $({\tilde\delta}, {\tilde\gamma}, \tilde{\delta}_0)$ 
for $(A,B,C)$ by inserting \scaling\ into \scIII:
\eqn\scIV{\eqalign{&2C^2- 3{\nu_2\over \nu_1} C -1=0\, , \cr
&B=\left({1-2C^2\over 1+4C^2}\right)\left({2D+\nu_2\over \nu_2}\right)\, , \cr
&A^2={{\nu_2^2}\over {(1-2C^2)(1+4C^2)(1+C^2)}} \left({{2D+\nu_2}\over {\nu_2}}\right)\, .  
 } }
For example, the first line is a quadratic equation that maps the parameter $\sinh{\tilde \gamma}\equiv C$ of the representative solution to the general ${\nu_1}\over {\nu_2}$:
\eqn\gt{\sinh{\tilde \gamma}={{1-\sqrt{1+{8\nu_1^2\over{9\nu_2^2}}}}\over{{4\nu_1\over{3\nu_2}}}}\, .}
For small ${\tilde \gamma}$ we have $\sinh{\tilde \gamma}\sim -{{\nu_1}\over{3\nu_2}}$. The subsequent  equations in \scIV\ determine  the coefficients $B$ and $A^2$ in terms of $C$,  $\nu_2$ and $D$.

%%%%%%%%%%%%%%%%%%
\subsec{The Subtracted Geometry}
We first consider the function ${\cal A}_{\rm red}(r)=L(r)$ given in \fe. We restrict to the special case corresponding to ``tilde" variables and then take the scaling limit \scaling. This does not change the functional form of the expression although of course the formulae for $({\tilde \nu}_1,\  {\tilde \nu}_2, \  {\tilde D})$ simplify to \scIII. The map \scII\ from the representative solution to the most general solution subsequently restores general ($\nu_1,\  \nu_2,\   D$) and all $\epsilon$'s cancel. The net effect is that ${\cal A}_{\rm red}$ in the subtracted geometry is the same as ${\cal A}_{\rm red}$ in the original geometry for completely general parameters. This is important because this function encodes the thermodynamics of the black hole. 

We next consider the warp factor variously referred to as $\Delta$ or $W^2$. This is a complicated function of parameters (reproduced in the appendix A) that is in particular quartic in the radial variable $r$. In this case the scaling simplifies the expression greatly. The
linear function in $r$ that remains 
\eqn\scW{{\tilde\Delta}=W^2_{\rm subtracted}\equiv \Delta= L^2 -(R-U)L_1^2\, ,}
is in fact precisely the subtracted expression \hc. In the scaling solution the $L_1$, $L_2$  and $L=L_1r+L_2$ are expressed in terms of 
the simplified functions \scIII\ but again the map \scII\ restores the formulae \haa\ that express $L_{1,2}$ through 
general $\nu_1,\nu_2,  D$ and $m$.

%%%%%%%%%%%%%%%%%%
\subsec{Matter for the Subtracted Geometry: Static Case}
The  scaling limit also allows for the determination of all the matter sources. We first give expressions for 
the static case ($a=0$) and then include rotation ($a\ne 0$).

In the scaling limit the three axio-scalars are equal and spatially constant: 
 \eqn\axioa{\eqalign{ 
&\chi= {{\tilde P}\over 2\tilde{Q}}~,
}}
where 
\eqn\ch{\eqalign{
{\tilde P}&= 2{\tilde m}\cosh{\tilde \delta}e^{\tilde{\delta}_0}\cosh{\tilde \gamma}\sinh{\tilde \gamma} 
{{1+4\sinh^2{\tilde \gamma}}\over {1-2\sinh^2{\tilde\gamma}}} ~,\cr 
{\tilde Q} &= 2{\tilde m}{\cosh^2{\tilde \delta}\cosh^2{\tilde \gamma}}{{1+4\sinh^2{\tilde \gamma}}\over {1-2\sinh^2{\tilde\gamma}}}\, .
}}
The variables $\tilde{P}$ and $\tilde{Q}$ are the large magnetic and electric charges of the dilute gas limit. They scale 
as $\epsilon^{-{2\over 3} }$ and $\epsilon^{-{4\over 3} }$, respectively. The value \axioa\ is identical to the attractor value for the extreme black hole noted in \gcf. The presence of a pseudo-scalars is characteristic of magnetic charges, 
even as $\chi$ in $\ch$ vanish in the strict scaling limit.

The bootstrap back to general charges gives the axion matter 
 \eqn\axio{\eqalign{ 
&\chi={{\sinh{\tilde \gamma}(mL_1+L_2)}\over {{\tilde Q}^2} }\,,
}}
in support of the general subtracted solution. The expression for the electric charge in terms of general parameters is of the form: 
\eqn\chr{
\tilde{Q}^3 = 2L_1 ( mL_1 + L_2)\epsilon^{-1} = (2m)^3 \left( 1 + {\nu_1^2\over\nu_2^2}\right)^2 \nu_2 (\nu_2 + 2D)\epsilon^{-1}~.
}
%
%This parameter effectively plays the role of central charge.
It carries an explicit $\epsilon$  dependence and it is thus formally infinite.  (The prefactor effectively plays the role of central charge.)

The remaining combinations are: 
\eqn\lc{mL_1+L_2=2m^2(1+{\nu_1^2\over \nu_2^2})(2D+\nu_2)\, ,} 
and 
$\sinh{\tilde \gamma}$ determined in terms of ${\nu_1}\over \nu_2$ through  \gt. 

The three parity-even scalars are also the same and equal to  
 \eqn\axio{\eqalign{ 
 &e^{\varphi_1} =e^{\varphi_2} =e^{\varphi_3} = {{\tilde Q}^2\over {\Delta}(a=0)}\, ,}}
where $\Delta$ is the subtracted geometry warp factor \scW, here evaluated in the static limit $U=0$. This $\Delta$ has radial dependence and so the scalars have a nontrivial flow from the confining asymptotic boundary to the black hole horizons. 

The Kaluza-Klein gauge potential ${\cal A}=A^4$ is turned on in the subtracted geometry. After the scaling limit and the bootstrap back to general charges we find:
\eqn\gauge{\eqalign{ 
&{\cal A} ={{{\tilde  Q}^3\left[ (L_2(2mL_1+L_2) -m^2{\nu_1^2\over \nu_2^2}L_1^2\right] } \over{2L_1(mL_1+L_2) \Delta(a=0)}}\,  dt\, .}}
The Kaluza-Klein  electric field acts effectively as being sourced by the localized electric charge :
\eqn\kkc{Q_{KK}=2m{{D(D+\nu_2)-{\textstyle{1\over 4}}\nu_1^2}\over{\left[(1+{\nu_1^2\over\nu_2^2})\nu_2(\nu_2+2D)\right]^2}}
\, .}
This charge is significant because it plays the role of conformal weight (or excitation energy) in the underlying CFT. 
In the four charge case: $\nu_1=0$,  $\nu_2=\Pi_c-\Pi_s$, and $D=\Pi_s$ where $\Pi_c=c_{\delta 0123}$ and $\Pi_s=s_{\delta 0123}$. 
Thus the effective Kaluza-Klein charge reduces to
\eqn\kkcf{Q_{KK}^0={{2m\Pi_s\Pi_c}\over {(\Pi_c^2-\Pi_s^2)^2}}\, .}
This is the result we reported in \CveticDN. 
 
The final matter field that is excited in the subtracted geometry is the gauge potential $A$, or more precisely three equal gauge potentials 
$ A=A^1=A^2=A^3$. To the leading order these fields vanish in the scaling limit but, as in the expansion of \bc\ for the magnetic charges, the second order is important (and more effort to compute). We find
\eqn\gauge{\eqalign{ 
A={{1\over {\tilde Q}}}\left(-r\, dt-{{ e}\over {\Delta(a=0)}}\, dt  +  p\sin\theta\, d\phi \right)\,  , }}
where 
\eqn\ep{\eqalign{
&{e}={\textstyle{1\over 2}}{m^2}\,{  L_1(mL_1+L_2 )}\sinh^2{\tilde \gamma }\left[\left({{2D+\nu_2}\over{\nu_2}}\right)^2-\left(1+{\nu_1^2\over \nu_2^2}\right)\right]\,, \cr
&{ p}={\textstyle{1\over 2}}\, m^2\cosh^2{\tilde \gamma}
(-{\nu_1\over 3})\left({{2D+\nu_2}\over \nu_2}\right)^2\, ,}}
and the expression for $\sinh{\tilde \gamma}$ is given in \gt. We simplified by gauging away a constant term $-dt$. The final gauge fields are dominated by a constant electric field. However,  since   $\sinh{\tilde \gamma}\ne 0$ (and thus $\nu_1\ne 0$), there are now also corrections due to the magnetic charge effects both for the $\phi$ and $t$ components.

We would also  like to comment on  the formally infinite  constant coefficient ${\tilde Q}$, that enters the expressions for the matter fields above.  Due to the symmetry of the Lagrangian (see appendix A)  we can rescale  the matter fields by the appropriate powers of ${\tilde Q}$:  three  equal dilatons by $e^{\varphi_i}$ by ${\tilde Q}^{-2}$; three equal axions $\chi_i$  by ${\tilde Q}^2$; the Kaluza-Klein  gauge potential ${\cal A}$  by ${\tilde Q}^{-3}$;  and the three equal remaining gauge potentials $A$ by ${\tilde Q}$. Such a scaling leaves the equations of motion invariant and thus, one can replace $\tilde Q$ in the matter fields above  with another  dimensionfull constant, say, a canonic factor  $2m$. 

%%%%%%%%%%%%%%%%%%
\subsec{Matter for the Subtracted Geometry: Rotation Included}
The addition of rotation makes computations more involved and results more elaborate. However, the logic leading from the scaling limit to the determination of sources supporting the general subtracted geometry with no restrictions on parameters is unchanged. We can therefore be brief. 

The three axio-scalars are equal and now take the following form (see Appendix B):
 \eqn\axiorot{\eqalign{ 
&\chi_1= \chi_2=\chi_3= { {a\sin\theta L_1 +\sinh{\tilde \gamma} (m   L_1+L_2)}\over {{\tilde Q}^2}}\,,
\cr
 &e^{\varphi_1} =e^{\varphi_2} =e^{\varphi_3} = {{\tilde Q}^2\over {\Delta}}\, .}}
where  ${\tilde Q}$ defined in \ch and $\Delta$ is the full subtracted geometry warp factor \scW .

The  Kaluza-Klein potential ${\cal A}\equiv A^4$ takes the form:
\eqn\gaugesr{
{\cal A} ={{{\tilde  Q}^3[(L_2(2mL_1+L_2)+(a^2\sin^2\theta -m^2{\nu_1^2\over \nu_2^2})L_1^2]}\over {2L_1(mL_1+L_2)\Delta}}\, dt  +{\tilde Q}^3 {a\cos^2\theta L_1\over  \Delta}\, d\phi\,,}
and the three equal potentials $A\equiv A^1=A^2=A^3$  take the  following form:
\eqn\gaugesrr{\eqalign{
A &=-{1\over {\tilde Q}}\left[r +a\sin\theta\sinh{\tilde \gamma}+{a^2\over 2m}\sin^2\theta\left( {\nu_2\over {2D+\nu_2}}\right)+{ f\over \Delta}\right]\, dt \cr
& +{1\over {\tilde Q}}\left[   p\sin\theta -a\cos^2\theta\, L_1\, \left(1  +{v\over \Delta}\right)\right]\, d\phi\,.}}
where $p$ is defined in \ep . The coefficients $f$ and $v$ are:
\eqn\fs{\eqalign{
{f}&=e+\, L_1(mL_1+L_2)\, \{am\sin\theta\, \sinh{\tilde \gamma }\left({\nu_2\over 2D+\nu_2}\right) \left[\left({{2D+\nu_2}\over{\nu_2}}\right)^2-\left(1+{\nu_1^2\over \nu_2^2}\right)\right] \cr
&+{\textstyle{1\over 2}}a^2\sin^2\theta\left[\cosh^2{\tilde\gamma}-\left(1+{\nu_1^2\over \nu_2^2}\right)\left({\nu_2\over 2D+\nu_2}\right)^2\right]\cr
&+{a^3\sin^3\theta\over m}\sinh{\tilde \gamma}\left({\nu_2\over 2D+\nu_2}\right)+{\textstyle{1\over 2}}{a^4\sin^4\theta\over m^2}\left({\nu_2\over 2D+\nu_2}\right)^2 \} \, ,\cr
v&=L_1(mL_1+L_2)\, \left[m\sinh^2{\tilde \gamma }\left({2D+\nu_2\over \nu_2}\right)+2\,a\sin\theta\, \sinh{\tilde \gamma}+{a^2\over m}\sin^2\theta\left({\nu_2\over 2D+\nu_2}\right)\right] \, ,}}
where $e$  is defined in \ep\ and  the expression for $\sinh{\tilde \gamma}$ is given in \gt.  Above, we have used the parameterization $U=a^2\cos^2\theta$. In the equal gauge potentials  $A$ we have gauged away a constant term $\propto dt$.
%Again, in the above expressions one can gauge away ${\tilde Q}$, and replace it, say, with a canonical value $2m$.  
For completeness, recall that:
\eqn\recall{\eqalign{&L_1=2m\left(1+{\nu_1^2\over \nu_2^2}\right )\nu_2\, , \quad L_2=4m^2\left(1+{\nu_1^2\over \nu_2^2}\right) D\,, \cr
 &L_1(mL_1+L_2) =4m^3\left(1+{\nu_1^2\over \nu_2^2}\right)^2(2D+\nu_2)\,\nu_2\, .} }

The gauge potentials have a straightforward radial  dependence also after inclusion of rotation: the Kaluza-Klein potential ${\cal A}$ dominated by the localized electric charge, and the gauge potentials $A$ by a constant electric field. However, due to the interplay of the additional magnetic charge  effects, associated with ${\tilde \gamma}\ne 0$,  and the non-zero rotational parameter $a$, there are additional polar angle dependent terms both with the $t$ and $\phi$ components, proportional to powers of $\sinh{\tilde \gamma}$ and/or powers of $a$.
Note again that the symmetry of the equations of motion allows one to replace $\tilde Q$  in the above expressions by  another finite  dimension full constant, say, $2m$.

While the results above were obtained by performing a scaling limit, it is expected that the subtracted geometry can also be obtained 
by acting with specific four Harrison transformations (with three boosts taken to infinity and the fourth one finite) on the original black hole. 
This procedure was previously carried out in the four charge case \refs{\CveticTR,\VirmaniKW,\CveticCJA}.

%%%%%%%%%%%%%%%%%%%%%%%%%%%%%%%%%%
\newsec{Connection to AdS$_3$}
In this section we lift the subtracted geometry to five dimensions and find a $S^2$ fibered over a BTZ base. The parameters of the effective BTZ give conformal weights and central charges that reproduce the entropy of the general 4D black hole. 

\subsec{5D Lift}
The metric \fa\ has apparent singularities where $G(r,\theta)=0$. The full metric is such that in fact all singularities
cancel. It is a useful first step to show this explicitly for the subtracted warp factor \hc. Mixing the notations introduced in \fb\ and \fc\ we find
\eqn\ga{\eqalign{
\Delta^{1\over 2} ds^2_4  & = -{R-U\over\Delta}  ( dt  + {LU\over a(R-U)} d\phi)^2+ {X\over R-U}\sin^2\theta d\phi^2
+  {dr^2\over X} + d\theta^2 \cr
& = -{R-U\over\Delta}  dt^2  - 2 {LU\over a\Delta} d\phi dt + {\sin^2\theta\over R-U} (R - {L^2 U\over \Delta}) d\phi^2
+  {dr^2\over X} + d\theta^2\cr
& = {dt^2\over L_1^2} -  {1\over\Delta} \left( {L^2\over L_1^2} dt^2 + 2a\sin^2\theta L dtd\phi+a^2\sin^4\theta L_1^2 d\phi^2 \right)
+ {dr^2\over X} + d\theta^2 + \sin^2\theta d\phi^2\cr
& = {dt^2\over L_1^2}  -  {1\over L^2_1\Delta} \left( Ldt + a\sin^2\theta L^2_1 d\phi\right)^2
+ {dr^2\over X} + d\theta^2 + \sin^2\theta d\phi^2~.
}}
We used the effective potential \hd\ in the form
\eqn\gb{
{\Delta - L^2\over R-U} = - L_1^2~,
}
repeatedly.

The second term in \ga\ appears to obstruct separability because it mixes $r$ and $\theta$ but, as we have already discussed, this is a false impression. We can make separability manifest by introducing the auxiliary gauge field
\eqn\gc{
{\cal B}= {1\over L_1\Delta} \left( Ldt + a\sin^2\theta L^2_1 d\phi\right)~,
}
and the auxiliary 5D geometry
\eqn\gd{\eqalign{
ds^2_5 &= \Delta (d\alpha + {\cal B})^2 + \Delta^{-{1\over 2}}ds^2_4\cr 
& = 
{1\over L_1^2} dt^2 
+ {dr^2\over X} + d\theta^2 + \sin^2\theta d\phi^2 + \Delta d\alpha^2 + 2d\alpha {1\over L_1} \left( Ldt + a\sin^2\theta L^2_1 d\phi\right)\cr
& = 
{1\over L_1^2} dt^2 
+ {dr^2\over X} + d\theta^2 + \sin^2\theta (d\phi + aL_1 d\alpha)^2 + (\Delta- a^2\sin^2\theta L_1^2) d\alpha^2 + 2{L\over L_1}  d\alpha dt \cr
& = 
{1\over L_1^2} (1 - {L^2\over \rho})dt^2 
+ {dr^2\over X} + \rho (d\alpha+ {L\over \rho L_1}  dt )^2+ d\theta^2 + \sin^2\theta (d\phi + aL_1 d\alpha)^2 \cr
& = 
- {X\over \rho}dt^2 
+ {dr^2\over X} + \rho (d\alpha+ {L\over \rho L_1}  dt )^2+ d\theta^2 + \sin^2\theta (d\phi + aL_1 d\alpha)^2 ~.
}}
The radial coordinate
\eqn\ge{
\rho = \Delta - L_1^2 a^2 \sin^2\theta~,
}
is independent of $\theta$. The metric \gd\ is manifestly separable. 

The potential ${\cal B}$ obtained here from separability agrees with the source ${\cal A}$ determined from the scaling limit and
given in \gauge. The expressions are not identical in that ${\cal B}$ asymptotes a constant term $[2L_1(mL_1+L_2)]^{-1}\, dt$ 
while ${\cal A}$ vanishes asymptotically. However, this amounts to a difference in gauge so the fields are physically equivalent.

\subsec{The Effective BTZ}
The final form \gd\ is locally AdS$_3\times S^2$. We can analyze its parameters using standard methods, as follows. 

The 5D geometry \gd\ was introduced such that the metric is a pure number: the mass dimension of $\Delta$ is $-4$ and so the coordinate $\alpha$ has mass dimension $2$. The periodicity of $\alpha$ is $2\pi R_\alpha$ where $R_\alpha$ has mass dimension $2$. In these conventions the radii of the AdS$_3$ and $S^2$ are pure numbers: $\ell_A=2$ and $l_S=1$. 

The $S^2$ is fibered over the AdS$_3$: the effective azimuthal angle in \gd\ is $\phi_{\rm eff} = \phi+aL_1\alpha$. Periodicity of $\phi_{\rm eff}$ is compatible with the periodicity of $\phi$ only if the period of $\alpha$ is such that 
\eqn\gf{
{1\over aL_1 R_\alpha} = {\nu_2\over 2ma (\nu_1^2 + \nu_2^2)R_\alpha} = {1\over 2G_4 J R_\alpha}= w. 
}
is an integer. Generally the $S^2$ in \gd\ will in fact be subject to $Z_w$ identifications. 

We can recast the 3D base of the 5D metric \gd\ in the standard BTZ form
\eqn\gga{
ds^2_{\rm BTZ} = - {(\rho^2_3 - \rho^2_+)(\rho^2_3 - \rho^2_-)\over \ell^2_A \rho^2_3}dt^2_3
+ { \ell^2_A \rho^2_3\over (\rho^2_3 - \rho^2_+)(\rho^2_3 - \rho^2_-)} d\rho^2_3 
+ \rho^2_3 (d\phi_3 + {\rho_+\rho_- \over \ell_A \rho^2_3} dt_3)^2~,
}
where the BTZ coordinates
\eqn\gh{\eqalign{
\rho^2_3 & = R^2_\alpha \left( (2L_1 (m L_1 + L_2) r + L_1^2 ( m^2 {\nu^2_1\over\nu^2_2} - a^2) + L^2_2 \right)~,\cr
t_3 & = {\ell_A\over R_\alpha }{1\over 2L_1 (m L_1 + L_2)} t~, \cr
\phi_3 & = {\alpha\over R_\alpha} + {t_3\over\ell_A}~.
}}
The term in $\phi_3$ that is proportional to $t_3$ is due to the constant in ${\cal B}$ mentioned below \ge. It can be gauged away as it was in our presentation of the source ${\cal A}$. 

The computation determines the mass and angular momentum of the effective BTZ black hole as
\eqn\gi{\eqalign{
M_3 & = {\rho^2_+ + \rho^2_-\over 8G_3\ell^2_A} 
= {m^2 R^2_\alpha \over 4G_3} ( 1 + {\nu^2_1\over\nu^2_2})^2 \left( 4m^2 D(D+\nu_2) + 2m^2 \nu^2_2 + m^2 \nu^2_1 - a^2 \nu_2^2\right)~, \cr
J_3 & = {\rho_+ \rho_-\over 4G_3\ell_A}= {m^2 R^2_\alpha \over 2G_3} 
( 1 + {\nu^2_1\over\nu^2_2})^2 \left( 4m^2 D(D+\nu_2) - m^2 \nu^2_1 + a^2 \nu_2^2\right)~.
}}

The effective gravitational coupling $G_3$ in 3D is related to the 4D Newton's constant $G_4$ by the comparison between reduction of 5D gravity on a sphere with radius $\ell_S=1$ and a circle with radius $2\pi R_\alpha$. Altogether the Brown-Henneaux formula for the central charge of AdS$_3$ with radius $l_A=2$ becomes
\eqn\gj{
k = {c\over 6}  = {\ell_A\over 4G_3} = {\ell_A\over 4} \cdot {4\pi \ell^2_S\over G_5} = {1\over R_\alpha G_4} = 2J w~.
}
In the final step we introduce the integral fibration period $w$ from \gf. The value $w=1$ gives the Kerr/CFT central charge $c=12J$. It is singled out as a manifest duality invariant.  

The physical parameters \gi\ correspond to the AdS$_3$ conformal weights
\eqn\gk{\eqalign{
h^{\rm irr}_L & = {M_3\ell_A + J_3\over 2}  = {R_\alpha\over G_4} ( F + I_4 )~, \cr
h^{\rm irr}_R & = {M_3\ell_A - J_3\over 2} = {R_\alpha\over G_4} (F - G_4^2 J^2) ~.
}}
The central charge \gj\ gives the Cardy formula
\eqn\gl{
S =  2\pi\sqrt{kh^{\rm irr}_L} + 2\pi\sqrt{kh^{\rm irr}_R} = 2\pi\sqrt{F+I_4} + 2\pi \sqrt{F- J^2} ~,
}
in units where $G_4=1$. This is the correct entropy of the 4D black hole in full generality. 

\newsec{Summary}
In summary, we have developed two distinct but related approaches to the analysis of the Chow-Comp{\`e}re black holes to the 4D STU-model. Taken together they show that much of the standard lore extends to the setting that includes both angular momentum and intrinsic spin. 

For clarity we take this opportunity to summarize in more detail the interconnection between the different parts of our work. 
One of the main lines of logic is:

\itemitem{---}
Start from the general black hole. 

\itemitem{---}
Tune parameters according to the dilute gas prescription.

\itemitem{---}
Simplify the entropy in this limit.

\itemitem{---}
Model the system as a CFT by reading off the effective conformal weights needed to 
account for the entropy. 

%%%%%%%%%
\medskip

\noindent
The other main line of logic is: 

\itemitem{---}
Start from the general black hole. 

\itemitem{---}
Exploit the structure of the wave equation to construct the auxiliary subtracted geometry, still with general charges. 

\itemitem{---}
Uplift the subtracted geometry to AdS$_3\times S^2$ in 5D. 

\itemitem{---}
Identify an effective BTZ black hole.

\itemitem{---}
Determine conformal weights from its mass and angular momentum. 

%%%%%%%%%%%%
\medskip

\noindent
The two main lines of logic are tied together by additional considerations. Some concern the conformal weights: 

\itemitem{---}
The conformal weights determined by the two lines of development are the same. This is a consistency check 
 (and also a nontrivial check on algebra.) 
 
\itemitem{---}
The relation between macroscopic charges and microscopic (quantized) charges requires due consideration of the moduli that are fixed in the dilute gas regime.

\itemitem{---}
The conformal weights expressed in terms of microscopic charges agree with values previously obtained from the lattice of $U(1)$ charges in the dual CFT. 

%%%%%%%%%%%%
\medskip

\noindent
There are also additional considerations that probe the full geometry:

\itemitem{---}
The subtracted geometry can be derived from the full solution as a scaling limit. 

\itemitem{---}
The scaling limit determines the sources supporting the affective AdS$_3\times S^2$ solution in 5D.

\medskip

\medskip

\noindent {\bf Acknowledgments:} \medskip \noindent
We thank David Chow, Geoffrey Comp{\`e}re, Gary Gibbons, Hong L\" u and Chris Pope for discussions. We thank the Michigan Center for Theoretical Physics for hospitality as this work was initiated. We also thank the Aspen Center for Physics, the Solvay Institute, and especially  the KITPC for hospitality. 
MC is supported by the DOE  Grant DOE-EY-76-02- 3071, the Fay R. and Eugene L. Langberg Endowed Chair, the Slovenian 
Research Agency (ARRS), and the Simons Foundation Fellowship. The work of FL was supported by  DoE grant DE-FG02-95ER40899.

\newsec{Appendix A: Lagrangian and Summary of the Chow-Comp\`ere Solution}
The Lagrangian density for the bosonic sector four-dimensional of the
 STU model (${\cal N}=2$ supergravity coupled to three vector supermultiplets) is
%%%%%
\eqn\lag{\eqalign{
{\cal L}_4 &= R\, {*{\bf 1}} - {1\over 2} {*d\varphi_i}\wedge d\varphi_i 
   - {1\over 2} e^{2\varphi_i}\, {*d\chi_i}\wedge d\chi_i - {1\over 2} e^{-\varphi_1}\,
( e^{\varphi_2+\varphi_3}\, {*   {\cal F}^1}\wedge  {\cal F}^1 \cr
 &+e^{\varphi_2-\varphi_3}\, {*  {\tilde {\cal F}}_2}\wedge    {\tilde {\cal F}}_2
  + e^{-\varphi_2 + \varphi_3}\, {*  {\tilde {\cal F}}_3 }\wedge    {\tilde {\cal F}}_3  + 
     e^{-\varphi_2 -\varphi_3}\, {* F^4}\wedge   F^4)\cr
&- \chi_1\, ( {\cal F}^1 \wedge  F^4+ {\tilde {\cal F}}_2 \wedge  {\tilde{\cal F}}_3 )
                 \,,
}}
%%%%%
where the index $i$ labelling the dilatons $\varphi_i$ and axions $\chi_i$
ranges over $1\le i \le 3$.  The field strengths $F^I=dA^I$  and their duals $d{\tilde F}_I=d{\tilde A}_I$ are related as:

%%%%%
\eqn\fp{\eqalign{
 {\cal F}^1 &= F^1 + \chi_2\,{\tilde F}_3 +  
    \chi_3\,  {\tilde F}_2 
     - \chi_2\, \chi_3\, F^4\,,\cr
 {\tilde {\cal  F}}_2 &= {\tilde F}_2 - \chi_2\,F^4\,,\cr
 {\tilde {\cal F}}_3 &={\tilde F}_3 - \chi_3\, F^4\,.\cr
}}
%%%%%
In the following we summarize the Chow-Comp\` ere black hole solution for reference. 
See \ChowTIA\ for further details.

The metric is:
%%%

\eqn\met{\eqalign{& ds^2  = - {{R - U}\over {W}} \left(d t + \omega_3 \right)^2
+ W \left( {d r^2\over R} + {d u^2\over U} + {{R U}\over {a^2 (R - U)} }\,  d\phi^2 \right) \, ,}}
where
\eqn\ww{\eqalign{
&W^2  = (R -  U)^2  + (2 N u + L)^2 + 2 (R - U) \left( 2 M r + V \right) \, , \cr
&\omega_3 =  {{2 N (u-n) R + U (L +2 N n)}\over {a (R - U)} }\, d \phi\, , \cr
&R(r)  = r^2 -2 m r + a^2 - n^2 , \qquad U(u) = a^2 - (u - n)^2\,  , }}
and
\eqn\LV{\eqalign{
&L(r)  = 2(-n \nu_1 + m \nu_2 ) r + 4(m^2 +n^2)D\, , \cr
&V(u)  = 2(n \mu_1 - m \mu_2)u +2 (m^2+n^2) C\, .}}
The standard choice for $U(u)=a^2\cos^2\theta$.

The potentials are
\eqn\munu{\eqalign{
\mu_1 & = 1 + \textstyle \sum_I [ {{1}\over {2}} (s_{\delta I}^2 + s_{\gamma I}^2) - s_{\delta I}^2 s_{\gamma I}^2 ] + {{1}\over{2}} \textstyle \sum_{I, J} s_{\delta I}^2 s_{\gamma J}^2 \, , \cr 
%%%
\mu_2 & = \textstyle \sum_I s_{\delta I} c_{\delta I} [ (s_{\gamma I} / c_{\gamma I}) c_{\gamma 1 2 3 4} - (c_{\gamma I} / s_{\gamma I}) s_{\gamma 1 2 3 4} ]\,  ,\cr 
%%%
\nu_1 & = \textstyle \sum_I s_{\gamma I} c_{\gamma I} [ (c_{\delta I} / s_{\delta I}) s_{\delta 1 2 3 4} - (s_{\delta I} / c_{\delta I}) c_{\delta 1 2 3 4} ] \,,  \cr
%%%
\nu_2 & = \iota - D\,  ,}}
%%%
%%%
where
\eqn\iotap{\eqalign{
%%%
\iota & = c_{\delta 1 2 3 4}c_{\gamma 1 2 3 4}+s_{\delta 1 2 3 4} s_{\gamma 1 2 3 4} 
%\cr & \quad 
+ \textstyle \sum_{I < J} c_{\delta 1 2 3 4} (s_{\delta I J} / c_{\delta I J}) (c_{\gamma I J} / s_{\gamma I J}) s_{\gamma 1 2 3 4}\,  , \cr
%%%
D & = c_{\delta 1 2 3 4}s_{\gamma 1 2 3 4} + s_{\delta 1 2 3 4}c_{\gamma 1 2 3 4} 
%\cr & \quad 
+ \textstyle \sum_{I < J} c_{\delta 1 2 3 4} (s_{\delta I J} / c_{\delta I J}) (s_{\gamma I J} / c_{\gamma I J}) c_{\gamma 1 2 3 4} \, ,\cr
C & = + 1 + \textstyle  \sum_I (s_{\delta I}^2 c_{\gamma I}^2 + s_{\gamma I}^2 c_{\delta I}^2) + \textstyle  \sum_{I < J} (s_{\delta I J}^2 + s_{\gamma I J}^2) + \textstyle  \sum_{I \neq J} s_{\delta I}^2 s_{\gamma J}^2 \cr
%%%
& \quad +  \textstyle \sum_I \sum_{J < K} (s_{\delta I}^2 s_{\gamma J K}^2 + s_{\gamma I}^2 s_{\delta J K}^2) 
%\cr & \quad 
+ 2 \textstyle  \sum_{I < J} [ s_{\delta 1 2 3 4} c_{\delta 1 2 3 4} (s_{\gamma I J}/c_{\delta I J}) (c_{\gamma I J}/s_{\delta I J}) \cr
%%%
& \quad + s_{\delta 1 2 3 4}^2 (s_{\gamma I J}^2 / s_{\delta I J}^2) + s_{\delta I J} s_{\gamma I J} c_{\delta I J} c_{\gamma I J} 
%%%
+ s_{\delta I J}^2 s_{\gamma I J}^2 ] 
%\cr &
- \nu_1^2 - \nu_2^2\, . }}
Here  $s_{\delta I} = \sinh \delta_I$, $c_{\delta I} = \cosh \delta_I$, $s_{\delta I \ldots J} = s_{\delta I} \ldots s_{\delta J}$, $c_{\delta I \ldots J} = c_{\delta I} \ldots c_{\delta J}$, and similarly for $\gamma$ instead of $\delta$. 

The solution depends on 11 independent parameters: the mass, NUT and rotation parameters ($m$, $n$, $a$);
and electric ($\delta_I$) and magnetic ($\gamma_I$) charge parameters. The physical charges correspond to these parameters in a complicated way. In particular, the mass $M$ and NUT charge $N$ are
%%%
\eqn\massnut{
M  = m \mu_1 + n \mu_2 \, , \quad N = m \nu_1 + n \nu_2 \,,}
where $\mu_1, \mu_2, \nu_1, \nu_2$ are  defined above as functions of $(\delta_I,\gamma_I)$.
For the  black hole solution we impose the zero Taub-NUT charge $N=0$, which constrains the bare Taub-NUT parameter 
\eqn\nutc{n=-{\nu_1\over\nu_2}m\, .}
%%%
The electric and magnetic charges are given as
%%%
\eqn\charges{
Q_I = 2{{\partial M}\over {\partial \delta_I} }\, , \quad  P^I  = -2 {{\partial N}\over {\partial \delta_I}}\,  .
}
The gauge fields take the  form:
%%%
\eqn\gaugef{
A^I  = W{{\partial }\over{\partial \delta_I} }W^{-1}(d\, t + \omega_3) \, ,}
which can be cast in an explicit form as:
\eqn\gaugeexp{
A^I  = \zeta^I (d\,t+ \omega_3) + A_{3}^I \,, }
where
%%%
\eqn\gauget{\eqalign{
W^2 \zeta^I & = {\textstyle{1\over 2}} {{\partial (W^2)}\over {\partial \delta_I}} 
 = (R - U) (Q_I r + {{\partial V}\over {\partial \delta_I}}) + (L +2 Nu) ( {{\partial L}\over {\partial \delta_I}} - P^I u ) \, , \cr
A_{3}^I & = {{P^I (u-n)}\over {a}} \, d\, \phi + {{U}\over {a (R - U)}} \left( P^I u - {{\partial L}\over {\partial \delta_I}}\right) \, d\, \phi\, .
}}
%%%%
The scalar fields are
%%%
\eqn\scalars{
\chi_i  = {{f_{i}}\over {r^2 + u^2 + g_i}} \,, \quad   {\rm e}^{\varphi_i}  = {{r^2 + u^2 + g_i}\over {W}} \, ,
}
%%%
where
%%%
\eqn\figi{\eqalign{
f_{i} & = 2 (m r + n u) \xi_{i 1} + 2 (m u - n r) \xi_{i 2} + 4 (m^2 + n^2) \xi_{i 3}\,  , \cr
g_{i} & = 2 (m r + n u) \eta_{i 1} +  2 (m u - n r) \eta_{i 2} + 4 (m^2 + n^2) \eta_{i 3}\, ,}}
and
\eqn\xieta{\eqalign{
%%%
\xi_{11} & = [ (s_{\delta 123}c_{\delta 4} - c_{\delta 123}s_{\delta 4} )s_{\gamma_1}c_{\gamma_1} + (1 \leftrightarrow 4) ] 
 \quad - ( (1,4) \leftrightarrow (2,3)) \, , \cr
%%%
\xi_{12} & = [{\textstyle{1\over 2}}(c_{\delta 23}s_{\gamma 14}+c_{\gamma 14} s_{\delta 23 })(c_{\delta 14}c_{\gamma 23} + s_{\gamma 23}s_{\delta 14}) \cr
%%%
& \quad + s_{\delta 1} s_{\gamma 4} c_{\delta 4} c_{\gamma 1} (s_{\delta 2} s_{\gamma 2} c_{\delta 3} c_{\gamma 3} + s_{\delta 3} s_{\gamma 3} c_{\delta 2} c_{\gamma 2}) 
%%%
+ (1 \leftrightarrow 4)] - ( (1,4) \leftrightarrow (2,3)) \, , \cr
%%%
\xi_{13} & = [( s_{\delta 1 3 4} c_{\delta 2} c_{\gamma 2}^2 + c_{\delta 1 3 4} s_{\delta 2} s_{\gamma 2}^2 ) s_{\gamma_3} c_{\gamma_3} + (2 \leftrightarrow 3) ] 
%%%
  - ( (1,4) \leftrightarrow (2,3)) \, , \cr
\eta_{1 1} & = s_{\delta 2}^2 + s_{\delta 3}^2 + s_{\gamma 1}^2 + s_{\gamma 4}^2 + (s_{\delta 2}^2 + s_{\delta 3}^2) (s_{\gamma 1}^2 + s_{\gamma 4}^2) 
%%%
 + (s_{\delta 2}^2 - s_{\delta 3}^2) (s_{\gamma 3}^2 - s_{\gamma 2}^2) \, , \cr
%%%
\eta_{1 2}&  = 2 s_{\delta 2} c_{\delta 2} (c_{\gamma 2} s_{\gamma 1 3 4} - s_{\gamma 2} c_{\gamma 1 3 4}) + (2 \leftrightarrow 3) \, , \cr
%%%
\eta_{1 3}&  = 2 s_{\delta 2 3} c_{\delta 2 3} (s_{\gamma 2 3} c_{\gamma 2 3} + s_{\gamma 1 4} c_{\gamma 1 4})+ s_{\delta 2 3}^2 (1 + \textstyle \sum_I s_{\gamma I}^2 ) \cr
%%%
 &\quad + (s_{\delta 2}^2 + s_{\delta 3}^2+2 s_{\delta 23}^2) (s_{\gamma 1 4}^2 + s_{\gamma 2 3}^2) 
+ s_{\delta 2}^2 s_{\gamma 2}^2 + s_{\delta 3}^2 s_{\gamma 3}^2 + s_{\gamma 1 4}^2 \, . }}
%%%

\newsec{Appendix B: Some Useful  Dilute Gas Expressions}

Here we display an explicit form of some quantities of the dilute gas black hole with  parameters  ${\tilde \delta_i}\equiv{\tilde\delta}, ( {\tilde \gamma_i}\equiv {\tilde\gamma}$ ($i=1,2,3$), and the fourth magnetic boost ${\tilde \gamma}_0=0$, and  with the hierarchy ${\tilde \delta}\gg ({\tilde \delta}_0, {\tilde \gamma})$. These expressions turn out to be useful in the derivation of the subtracted geometry as a scaling limit.

Specifically in this case :
\eqn\scalingt{\eqalign{{\tilde \nu}_1&\sim -3\cosh{\tilde \gamma}\sinh{\tilde \gamma}\exp(-{\tilde \delta}_0)\cosh ^3{\tilde \delta},  \cr
{\tilde \nu}_2&\sim (\cosh^2{\tilde \gamma}-3\sinh^2{\tilde \gamma})\exp(-{\tilde \delta}_0)\cosh^3{\tilde \delta}, \cr   {\tilde D}& \sim (\cosh^2{\tilde \gamma}\sinh{\tilde \delta_0}+3\sinh^2{\tilde \gamma\cosh{\tilde \delta_0})\cosh{\tilde \gamma}}\cosh^3{\tilde \delta}\,. } }
and ${\tilde \mu}_1$ and ${\tilde \mu}_2$ are sub-leading to  ${\tilde \nu}_{1,2}$, since they scale with $\cosh ^2{\tilde\delta}$.
%
%or equivalently
%
%\eqn\scpar{\eqalign{ 
%&{{-3 \sinh{\tilde \gamma}}\over {1-2\sinh^2{\tilde \gamma}}}= {\nu_1\over \nu_2} \, , \cr
%&  e^{{\tilde \delta}_0}{{1+4\sinh^2{\tilde \gamma}}\over {1-2\sinh^2{\tilde \gamma}}}={2{ D}+\nu_2\over \nu_2}\, , \cr
%& {{(1+4\sinh^2{\tilde \gamma})\cosh^2{\tilde \gamma}}\over {(1-2\sinh^2{\tilde \gamma})^2}}=1+{\nu_1^2\over \nu_2^2}\,  \, .
%}}This results in expressions:
 \eqn\scaling {\eqalign{ 
{{\tilde \nu}_1\over {\tilde \nu}_2}&\sim{{-3 \sinh{\tilde \gamma}}\over {1-2\sinh^2{\tilde \gamma}}}\, , \cr
{2{ \tilde D}+{\tilde \nu_2}\over {\tilde \nu_2}}& \sim e^{{\tilde \delta}_0}{{1+4\sinh^2{\tilde \gamma}}\over {1-2\sinh^2{\tilde \gamma}}}\, , \cr
1+{{\tilde \nu_1}^2\over{\tilde  \nu}_2^2}& \sim{{(1+4\sinh^2{\tilde \gamma})\cosh^2{\tilde \gamma}}\over {(1-2\sinh^2{\tilde \gamma})^2}}\,  \, ,
}}
Furthermore:
\eqn\xietatilde{\eqalign{&\xi_{i1}\sim -{\tilde \nu}_1\, , \quad \xi_{i2}\sim {\tilde \nu}_2, \, ,\cr
&\xi_{i3}\sim{\textstyle{1\over 2}}({\tilde \nu}_1 +{\tilde \chi}) \, , \quad {\tilde \chi} \equiv 2 \cosh^3{\tilde \delta}\sinh{\tilde \gamma}\cosh{\tilde \gamma}(1+\sinh^2{\tilde \gamma})\, , \cr
&\eta_{i3}\sim\cosh^4{\tilde \delta} \cosh^2{\tilde \gamma}(1+4\sinh^2{\tilde \gamma})\, ,}}
and $\eta_{i1}$ and $\eta_{i2}$ are sub-leading to $\eta_{i3}$, since they scale with $\cosh^2{\tilde \delta}$.
These properties ensure that  axions take the following form:
\eqn\axiont{\chi_1\sim \chi_2\sim \chi_3\sim { {{\tilde a}\sin\theta {\tilde L}_1 +\sinh{\tilde \gamma} ({\tilde L}_2+{\tilde m} {\tilde L_1})}\over {{\tilde Q}^2}}\,,}
where
\eqn\cht{{\tilde Q}= 2{\tilde m}{{\cosh^2{\tilde \delta}\cosh^2{\tilde \gamma}(1+4\sinh^2{\tilde \gamma})}\over {(1-2\sinh^2{\tilde\gamma})}}\, ,}
is the large electric charge of the dilute gas. 

It is straightforward to see that the warp factor $W^2$ takes the form:
\eqn\warpf{W^2\sim {\tilde L}^2-({\tilde R}-{\tilde U}){\tilde L}_1^2\equiv {\tilde \Delta}\, ,}
and the dilatons are of the form:
\eqn\dilatonst{e^{\varphi_1} =e^{\varphi_2} =e^{\varphi_3} = {{\tilde Q}^2\over {\tilde \Delta}}\, .}

As for the gauge potentials,  one has to employ the formulae \gaugeexp and \gauget. While the calculation for the Kaluza-Klein potential ${\cal A}\equiv A^4$  is straightforward, a detailed care is needed to calculate the three equal gauge potentials $A\equiv A^1=A^2=A^3$, since the calculation  requires a subleading expansion in terms of $e^{\tilde \delta}$ parameter. The final result in  the scaling limit is
 presented in section 6.

\listrefs

\end